\documentclass[reprint,superscriptaddress,amsmath,amssymb,aps,prb,longbibliography]{revtex4-2}
\usepackage{amsmath, amsfonts}
\usepackage{float}
\usepackage{graphicx} 
\usepackage{dcolumn} 
\usepackage{bm} 
\usepackage[dvipsnames]{xcolor}
\usepackage[colorlinks=true,linkcolor=blue,citecolor=blue,urlcolor=black]{hyperref} 

\begin{document}

\title{Electronic States in One-Dimensional Helical Crystals: \\ General Properties and Application to InSeI}

\author{Jiaming Hu}
 \affiliation{School of Materials Science and Engineering, Zhejiang University, Hangzhou 310027, China}
 \affiliation{School of Engineering, Westlake University, Hangzhou 310030, China}
 \affiliation{Institute of Advanced Technology, Westlake Institute for Advanced Study, Hangzhou 310024, China}

\author{Shu Zhao}
\affiliation{School of Engineering, Westlake University, Hangzhou 310030, China}
\affiliation{Institute of Advanced Technology, Westlake Institute for Advanced Study, Hangzhou 310024, China}

\author{Wenbin Li}
\email{liwenbin@westlake.edu.cn}
\affiliation{School of Engineering, Westlake University, Hangzhou 310030, China}
\affiliation{Institute of Advanced Technology, Westlake Institute for Advanced Study, Hangzhou 310024, China}

\author{Hua Wang}
\email{daodaohw@zju.edu.cn}
\affiliation{Center for Quantum Matter, Zhejiang University, Hangzhou 310058, China}
\affiliation{ZJU-Hangzhou Global Scientific and Technological Innovation Center, School of Physics, Zhejiang University, Hangzhou 311215, China}

\date{\today}

\begin{abstract}
In this article, we systematically explore several key properties of electronic states in one-dimensional (1D) helical crystals, including the inheritance of orbital angular momentum (OAM) from local atomic orbitals to the entire helical structure, the helical momentum and the emergence of helical-induced spin-orbit coupling (hSOC). We then apply this comprehensive theoretical framework to elucidate the electronic structure of the 1D helical crystal InSeI. Our analysis reveals the influence of hSOC, evident in spin-mixing energy gaps within the electronic band structure, as calculated through density functional theory. Utilizing a combination of tight-binding modeling and first-principles calculations, we ascertain the spin-polarized electric response and the chiral-switchable second-order photocurrent response of InSeI, characterized as the Landauer-Buttiker ballistic transport and shift current response, respectively. The results highlight the potential of 1D InSeI for applications in spintronics and optoelectronics. The overarching theoretical framework established in this work will prove invaluable for the investigation of other helical electronic systems.

\begin{description}
\item[Keywords]
helical system; orbital angular momentum; spin-orbital coupling; spintronics; InSeI
\end{description}
\end{abstract}

\maketitle


\section{\label{sec:introduction}Introduction}

\textcolor{blue}{
Recently, helical materials, i.e. materials with atoms or groups of atoms arranged in a helical structure, similar to the well-known DNA double helix and chiral nanotubes, have attracted considerable interest due to their promising applications in electronic, optoelectronic, thermoelectric, and spintronic fields~\cite{review_chiral_spintronics, yan2023structural}. A pivotal aspect of the physics involved is the presence of chirality in helical structures, which leads to the celebrated chiral-induced spin-selectivity (CISS) effect~\cite{review_CISS}. Past experiments have proved significant CISS in electron transmission through double-stranded DNA on gold~\cite{ray2006chirality,gohler2011spin}, and reported chiral-involved spin response in chiral molecules~\cite{kumar2017chirality}. Recent studies on carbon helices~\cite{naskar2023chiral}, circular helical molecules~\cite{chen2023spin}, chiral graphene sheets~\cite{firouzeh2023chirality}, isolated chiral molecules~\cite{eckvahl2023direct} and semiconductor-chiral molecule heterojunctions~\cite{liu2020linear} also verify the existence of CISS. It is also reported that chiral materials could have CISS and spin polarization effects on a number of chemical reactions controlled by the electron’s spin~\cite{naaman2020chiral}. }

\textcolor{blue}{
The understanding of the fundamental mechanism of CISS has evolved over time. Initially, Bardarson's work proved that, in a two-terminal system composed of isolated transmission channels, spin polarization is prohibited if the time reversal symmetry is preserved~\cite{bardarson2008proof}. Later, under the framework of non-equilibrium Green's function, Utsumi and collaborators found that the mixed multi-channel electron transmission can support CISS without the breaking of time reversal symmetry~\cite{utsumi2020spin}. In their work, spin-orbit coupling (SOC) is addressed as the key effect of mixing different transmission channels, which was later confirmed by a series of transport studies~\cite{model_ballistic, model_DNA, DNA_TB_model,medina2012chiral,yeganeh2009chiral,michaeli2019origin, liu2023spin,wolf2022unusual}. Subsequently, Liu et al. pointed out the essence of multi-channel mixing as the orbital angular momentum (OAM) polarization~\cite{DFT_TB_orbitalpolarization}, which is discussed based on a tight-binding (TB) model of DNA-like materials with the OAM quantum number explicitly considered. As another view, an anisotropic-wire model for chiral molecules is reported in analyzing helical systems~\cite{ghazaryan2020analytic}. Experiments also suggest the source of chiral-involved spin response in chiral molecules as the charge reorganization accompanied by a polarization of the spins~\cite{kumar2017chirality}. A recent \textit{ab initio} study proves that the non-equilibrium spin polarization could be a key component in understanding the CISS mechanism~\cite{naskar2023chiral}. Studies for more realistic cases with finite temperature step further to experiments and applications~\cite{fay2021origin,chiesa2023chirality}. Additionally, there are also studies proposing interface-dominated~\cite{alwan2021spinterface}, polaron-induced~\cite{klein2023giant} and molecular vibration-induced~\cite{fransson2021charge} mechanisms. }


The aforementioned pioneering works have shed light on the unique electronic states in helical systems from different angles and demonstrated that the electronic states in helical systems exhibit distinct properties. However, there is still a lack of a comprehensive and systematic description of the electronic states in helical materials, which is expected to establish a complete relationship between the helical structure and all the important elements mentioned above (i.e. OAM, SOC, and CISS). Most previous research has focused on the effect of chirality on electron transport, with few studies focusing on the more fundamental influence of structural helicity on the intrinsic electronic states. Here it is worth mentioning the work of Hu and co-authors, in which a general expression for screw-symmetry induced SOC was obtained from symmetry group analysis and successfully applied to describe the spin texture of dislocations in conventional semiconductors~\cite{theoretical_screwSOC}. Similar lines of research that can directly relate the helical-induced SOC to the intrinsic electronic and spin structure of helical materials are highly desirable.  


Furthermore, due to the intricate geometry of helical materials, conventional electronic-structure modeling methods, such as the celebrated TB method, become complicated in describing realistic helical materials~\cite{DNA_TB_model,huertas2006spin}. As a consequence, it remains unclear how OAM can be explicitly included in the TB modeling, necessitating the development of adaptive methods for modeling the electronic structure of helical materials. 

Another important consideration pertains to the categorization of helical materials. In the early stages of research, the majority of helical materials discovered fell under the category of organic molecules or chiral nanotubes, and consequently most investigations have focused on these types of systems~\cite{model_ballistic, model_DNA, DNA_TB_model, medina2012chiral,yeganeh2009chiral,fransson2021charge,alwan2021spinterface, huertas2006spin,nitti2017chiral,dubi2022spinterface}. However, as synthesis technologies have advanced, it has become possible to create a variety of inorganic crystal materials with helical structures that exhibit exceptional optical, electronic, and mechanical properties~\cite{review_inorganichelices, SnIP, LiP}. Specifically, the quasi one-dimensional (1D) crystal indium selenium iodide (InSeI), which consists of 1D helical chains weakly bound together through van der Waals (vdW) forces, has recently been reported both theoretically~\cite{InSeI_DFT,Zhao2023, Zhou2023} and experimentally~\cite{InSeI_exp,Cordova2024}. These 1D InSeI helical chains possess a screw axis as the only non-translational symmetry element, with the space groups $P4_1$ and $P4_3$ for right-handed and left-handed chains, respectively, and are thus expected to become an archetypal 1D solid-state helical crystal. The achievement of single chain exfoliation in experiments~\cite{InSeI_exp} makes 1D InSeI more practical in future applications. Besides, since InSeI is composed of relatively heavy elements compared to organic materials, its SOC effect is also stronger than that in organic materials. However, only a limited number of theoretical analyses have addressed the properties of helical InSeI~\cite{Zhao2023,Zhou2023}, thereby impeding the comprehensive exploration of its potential applications.

Therefore, in this paper, we present a systematic discussion of the electronic states in \textcolor{blue}{isolated 1D helical crystals that exhibit periodicity in the axial direction (also referred to as helical materials or helical systems for simplicity in the following discussion). By ``1D crystals'', we refer to materials that are infinitely long along the axial direction with translational symmetry, while the term "helical" implies that the axial direction also corresponds to a screw axis. The term ``isolated'' indicates that we omit inter-chain interactions, a reasonable approximation in many realistic scenarios where 1D chains in quasi-1D bulk materials are bound by weak vdW interactions~\cite{InSeI_exp} within bulk phases, ensuring that the low-energy electronic states remain relatively unaffected. The physics of multi-chain systems, while also of interest, will be explored in future works.} 

The content of the paper is as follows. In Sec.~\ref{sec:general} we establish a general TB modeling procedure for isolated 1D crystal with a helical chain structure, where the interplay between helical structure and OAM is naturally captured as the OAM inheritance, helical momentum and helical-induced SOC (hSOC). The theory is subsequently applied to the analysis of the electronic structure and various properties of helical 1D InSeI in Sec~\ref{sec:app in InSeI}. The comparison between TB modeling and density functional theory (DFT) calculations supports the validity of our general theory, as shown by the OAM inheritance and hSOC in 1D InSeI in Sec.~\ref{subsec:OAM_InSeI} and Sec.~\ref{subsec:hSOC_InSeI}, respectively. In Sec.~\ref{subsec:helical-CISS} and Sec.~\ref{subsec:helical-SC}, we further uncover the properties of 1D InSeI that include CISS enabled by helical-induced spin-orbit coupling (hSOC) and chirality-dependent shift current response. We summarize our findings and draw our conclusion in Sec.~\ref{sec:conclusion}. Together, our work is expected to offer a comprehensive and systematic framework for modeling and understanding electronic states in 1D helical crystals.

\section{General case}\label{sec:general}

\subsection{ The influence of helical structure to OAM }\label{sec:modeling_TB_basis}

\begin{figure}
    \centering
    \includegraphics[width=0.35 \textwidth]{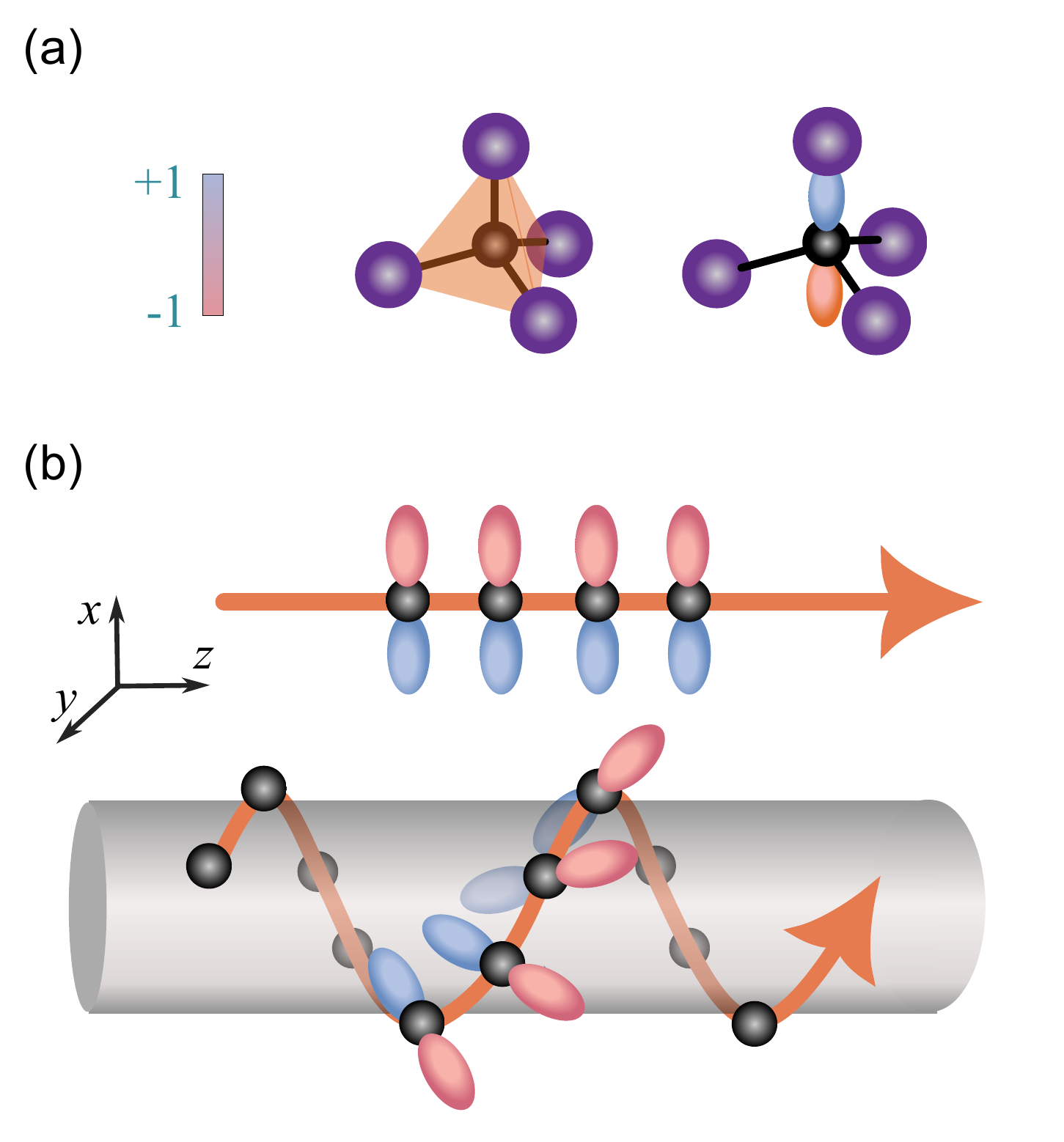}
    \caption{\label{fig:0}
    \textbf{Comparison between non-helical and helical system.} (a) Schematic of tetrahedron-type local chemical environment, and a $p$-type profile function (with phase labelled by different colors). (b) Schematic of identically polarized profile functions in a non-helical chain system, and tilted profile functions in a helical chain system. The purple atoms are omitted for simplicity. }
\end{figure}

Firstly we would like to briefly review the basic ideas of local chemical environment and TB modeling. In a bulk crystal, electrons are no longer isolated but interact with each other. In a perturbative view, however, the low-energy physics of the equilibrium electronic states (eigen-states) are mainly determined by local interactions, namely the interactions between orbitals at neighboring atomic sites, which makes the local chemical environment (also known as the ligand field in coordination chemistry) of great importance. In a TB model, one assigns the same set of localized wavefunctions as the unperturbed basis at the same types of atomic sites, and then adds interactions between the basis~\cite{book_solids}. Here, the same atomic sites refer to sites with the identical local chemical environment that includes the atomic type and bonding configuration. With the consideration of translational symmetry, the TB basis wavefunctions are written as 
\begin{equation}\label{eq:Psi_general}
    \Psi_{\lambda\gamma\pmb{k}}(\pmb{r}) = {\sum\limits_j}e^{i\pmb{k}{\cdot}\pmb{R}_j} u_{\lambda}(\pmb{r} -\pmb{r}_{\gamma} - \pmb{R}_j),
\end{equation}
where $\pmb{R}_j$ and $\pmb{r}_{\gamma}$ are the position vetors of the $j$-th crystal unit cell and the basis vector of the $\gamma$-th atomic site, respectively, while $\lambda$ indicates the type of localized orbitals. $u_{\lambda}$, the profile functions,  can be taken as maximally localized Wannier functions~\cite{Wannier}, which are mainly determined by the local chemical environment. As shown in Fig.~\ref{fig:0}(a), one atomic $p$-type profile function with its phase labelled in different colors is centered at the black atomic site, which is surrounded by a tetrahedron-type local chemical environment (coordinated with four nearest neighboring atoms in purple). 

In traditional non-helical systems, the local chemical environment of the same type atomic sites could overlap with each other through pure translations. That is, the same type of atomic sites point to the identical directions, so do the profile functions. In this case, Eq.~\ref{eq:Psi_general} satisfies the TB requirement that $u_{\lambda}$ is able to be sufficiently localized. However, in helical materials (or more generally, in curved systems including nanotubes), the local chemical environment of the same type of atomic sites may have different pointing directions, i.e. be \textit{tilted} to each other. Such tilting of the local chemical environment also rotates the profile function since the profile function must fit the configuration of the local chemical environment. As is illustrated in Fig.~\ref{fig:0}(b), with only the black atomic sites plotted for simplicity, the profile functions are all polarized in the same direction in the non-helical chain system, but rotates in the helical chain system due to the tilted local chemical environment.

This tilting effect is important because the directional polarization of the profile function reflects the orbital angular momentum (OAM). For example, the atomic $s$ orbital has no OAM (with zero OAM quantum number), resulting in a spherically symmetric profile without directional polarization. In contrast, atomic $p,d,f$ orbitals with non-zero OAM exhibit polarized angular distributions to specific directions~\cite{atkins2011molecular}. Consequently, the tilting of the chemical environment becomes relevant for states derived from these orbitals. It should be emphasized that although the tilting effect exists in both tubular and helical systems, it is an isolated degree of freedom in the former (rotational symmetry) but composited with a translational order in the latter (helical symmetry), implying the coupling between the OAM and the linear momentum in the helical systems, which will be discussed later.

If we still adopt Eq.~\ref{eq:Psi_general} as the TB basis in the helical systems, the tilting effect causes the issue that although many atomic sites belong to the same type (i.e. sharing the same local chemical environment), we still could not represent them using one TB basis since their profile functions have a global tilting order. A common way to fix this problem is to simply regard those tilted sites as belonging to different types and assign individual TB basis to them, which leads to an enlarged TB modelling basis. The fundamental origin of this issue is that the traditional TB method with crystal unit cell and the basis defined in Eq.~\ref{eq:Psi_general} does not incorporate the helical symmetry (screw symmetry) and thus leads to redundant degrees of freedom. Consequently, based on all the analysis above, it is expected that an OAM-corrected TB method would be more suitable for modeling helical systems.


\subsection{Helical momentum}\label{sec:modeling_helical_momentum}

To comprehensively investigate the impact of helical-induced tilting of chemical environment on the electronic states, and to develop an appropriate TB modeling approach for helical systems, let us consider a simplified scenario of a 1D crystal with a helical chain structure. This system remains invariant under a $\zeta$-fold screw symmetry operation $S_z=R_z[{\Delta}\varphi]{\cdot}T_z[a_0]$ operating along the screw axis ($z$-axis), where $R_z[{\Delta}\varphi]$ is the rotation of ${\Delta}\varphi=2{\pi}/\zeta$ and $T_z[a_0]$ is the translation of distance $a_0$. The crystal unit cell can be evenly separated into $\zeta$ segments along the screw axis, which are identical in internal structure (local chemical environment) but pointing to different directions, being tilted by ${\pm}2{\pi}/\zeta$ along the screw axis relative to their nearest neighbors. These segments are defined as \textit{helical unit cells}. Here $\zeta$ must be an integer because for a helical crystal, the translational symmetry and the screw symmetry should be satisfied simultaneously so that the length of helical pitch $\zeta{a_0}$ should equal to that of the crystal unit cell. That is, in a 1D crystal with $\zeta$-fold screw symmetry, a crystal unit cell contains $\zeta$ helical unit cells.

\textcolor{blue}{
\begin{figure}
    \centering
    \includegraphics[width=0.45 \textwidth]{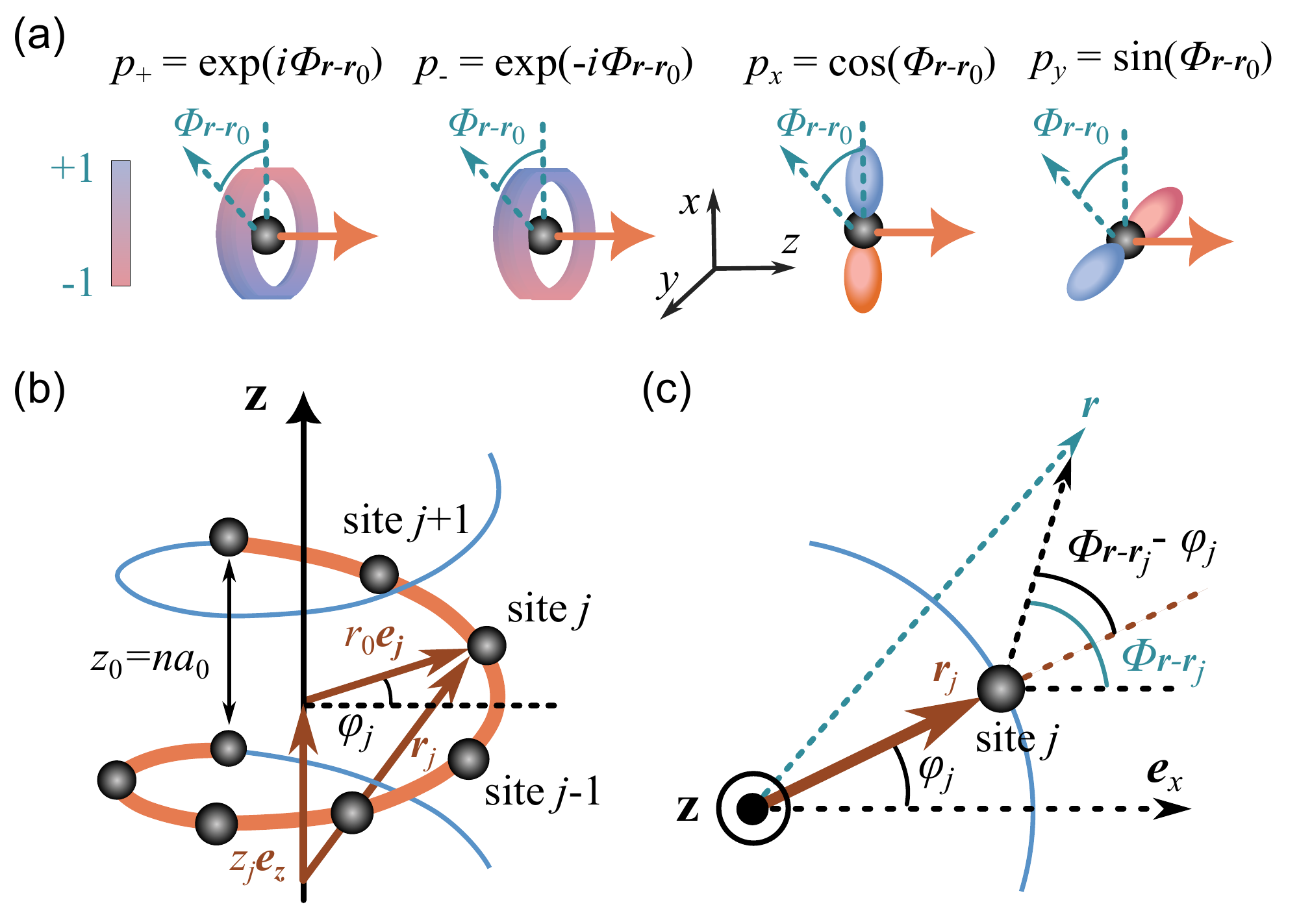}
    \caption{\label{fig:01} \textbf{Atomic orbital angular momentum (OAM) and helical geometry.} \textcolor{blue}{ (a) Schematic of atomic $p$-orbitals with color-coded phases. (b) Schematic of helical unit cells (with only one atomic site per helical unit cell). $\pmb{r}_j=z_j\pmb{e}_z+r_0\pmb{e}_j$ is the position of $j$-th helical unit cell with only one atomic site plotted, where $\pmb{e}_z$ and $\pmb{e}_j$ are the unitvectors of $z$-direction and radius direction respectively. (c) Schematic of $\Phi_{\pmb{r}-\pmb{r}_j}$ as the azimuthal angle of $\pmb{r}-\pmb{r}_j$ for arbitrary position vector $\pmb{r}$. Both $\varphi_j$ and $\Phi_{\pmb{r}-\pmb{r}_j}$ are measured from $\pmb{e}_x$. The brown dashed line is the extension of $\pmb{r}_j$. It is clearly found that the local angular coordinate should be $\Phi_{\pmb{r}-\pmb{r}_j}-\varphi_j$, which is measured fron the local reference direction $\pmb{r}_j$.} }
\end{figure}}

\textcolor{blue}{
In principle, there could be multiple atomic sites inside one helical unit cell. However, as we mentioned above, the distinguishing influence of helical structure that we focus on to the electronic states is the tilting of local chemical environment in a helical order (screw symmetry). According to our definition of helical unit cell, only the atomic sites in different helical unit cells are influenced by this helical-induced tilting effect. Therefore, for simplicity, we focus on the inter-cell part by considering only one atomic site inside each helical unit cell, so that the helical unit cell and the atomic site share the same label $j$ and position $\pmb{r}_j$. The extension to the multi-site case is straightforward and is also discussed in Sec.~\ref{sec:modeling_TBsummary}. As illustrated in Fig.~\ref{fig:01}(b), the location of $j$-th helical unit cell is
\begin{equation}\label{eq:r_j}
    \pmb{r}_j=z_j\pmb{e}_z+r_0\pmb{e}_j,
\end{equation}
where $\pmb{e}_z, \pmb{e}_j$ are the unit vectors along $z$ and radial direction, respectively, and $r_0$ is the radius of the helix. The azimuthal angle of $\pmb{e}_{j}$ is $\varphi_j$ with $\varphi_{j{\pm}1}-\varphi_j={\pm}2{\pi}/\zeta$. }

\textcolor{blue}{
Recall that with local spherical coordinates $\pmb{r}=(\pmb{r}|,\theta_,\Phi)$ of atom centered at the position $\pmb{r}_0$, the atomic orbital basis can be written as 
\begin{equation}\label{eq:atomic_nl}
    \psi_{nlm}(\pmb{r}-\pmb{r}_0) = A_{nl}(\pmb{r}-\pmb{r}_0)e^{im{\Phi}_{\pmb{r}-\pmb{r}_0}},
\end{equation}
where $A_{nl}$ is the modulus part with the main quantum number $n$ and angular quantum number $l$, while $m$ represents the OAM quantum number (also called the magnetic quantum number). Together with ${\Phi}_{\pmb{r}-\pmb{r}_0}$, $m$ determines the phase of $\psi_{nlm}$, which is the azimuthal angle of the vector ${\pmb{r}-\pmb{r}_0}$. For example, for $p$ orbitals we have~\cite{atkins2011molecular} 
\begin{equation}\label{eq:p_pm1_orbitals}
    \begin{aligned}
        |0{\rangle} 
        &=
        \psi_{n10}(\pmb{r}-\pmb{r}_0) 
        = 
        \sqrt{2}R_{n1}(|\pmb{r}-\pmb{r}_0|)
        \cos{\theta_{\pmb{r}-\pmb{r}_0}}, \\
        |{\pm1}{\rangle} 
        &=
        \psi_{n1,{\pm}1}(\pmb{r}-\pmb{r}_0)
        \\
        &=
        -R_{n1}(|\pmb{r}-\pmb{r}_0|)\sin{\theta_{\pmb{r}-\pmb{r}_0}}e^{{\pm}i{\Phi}_{\pmb{r}-\pmb{r}_0}},
    \end{aligned} 
\end{equation}
where $R_{n1}(\rho)$ is the radial part. A clear $e^{{\pm}i{\Phi}_{\pmb{r}-\pmb{r}_0}}$ term can be observed, which means that the phase of $\psi_{nl}$ changes as the local azimuthal angle changes, as schematically shown in Fig.~\ref{fig:01}(a). The linear combinations of the $p$ orbitals in a symmetric form leads to real $p_x$, $p_y$, and $p_z$ orbitals pointing toward $x,y,z$ directions, respectively:
\begin{equation}\label{eq:p_xyz_orbitals}
    \begin{aligned}
        p_x(\pmb{r}-\pmb{r}_0) &= \frac{1}{\sqrt{2}}(|-1{\rangle}+|+1{\rangle}) \ \propto \ \cos({\Phi}_{\pmb{r}-\pmb{r}_0}), \\
        p_y(\pmb{r}-\pmb{r}_0) &= \frac{i}{\sqrt{2}}
        (|-1{\rangle}-|{+1}{\rangle}) \ \propto \ \sin({\Phi}_{\pmb{r}-\pmb{r}_0}), \\
        p_z(\pmb{r}-\pmb{r}_0) &= |0{\rangle} .
    \end{aligned} 
\end{equation}}

\textcolor{blue}{
In a helical system, if we build TB model with helical unit cells as the fundamental structural units, then it is not appropriate to directly employ $\psi_{nlm}$ in the form shown in Eq.~\ref{eq:atomic_nl} as the profile function in the TB basis wavefunction of  Eq.~\ref{eq:Psi_general} as
\begin{equation}\label{eq:Psi_novarphi}
    \begin{aligned}
        \Psi_{nl,km}(\pmb{r})
        &=
        {\sum\limits_j}e^{ikz_j}A_{nl}(\pmb{r}-\pmb{r}_j)e^{im{\Phi_{\pmb{r}-\pmb{r}_j}}}.
    \end{aligned}
\end{equation}
The challenge arises from the fact that the reference direction of the local rotational coordinate $\Phi_{\pmb{r}-\pmb{r}_j}$, highlighted as $r_0\pmb{e}_j$ in Fig.~\ref{fig:01}(b), differs at different sites, which makes $\Phi_{\pmb{r}-\pmb{r}_j}$ not well-defined globally. To directly illustrate this point, we consider the matrix element $H_{12}$ of crystal Hamiltonian $H(\pmb{r})$ (i.e. hopping parameter of TB basis) between $\Psi_{n_1l_1,k_1m_1}(\pmb{r})$ and $\Psi_{n_2l_2,k_2m_2}(\pmb{r})$ 
\begin{equation}\label{eq:H_12_1}
    \begin{aligned}
        &H_{12}
        = 
        {\int}d\pmb{r}
        \Psi^*_{n_1l_1,k_1m_1}(\pmb{r})
        H(\pmb{r})
        \Psi_{n_2l_2,k_2m_2}(\pmb{r})
        \\
        &=
        {\sum\limits_{jj'}}
        e^{i(k_2z_{j'}-k_1z_j)}
        {\int}d\pmb{r}
        A_{n_1l_1}(\pmb{r}-\pmb{r}_j)
        H(\pmb{r})
        A_{n_2l_2}(\pmb{r}-\pmb{r}_{j'})
        \\
        &{\times}
        e^{i[m_2{\Phi_{\pmb{r}-\pmb{r}_{j'}}}-m_1{\Phi_{\pmb{r}-\pmb{r}_j}}]}.
    \end{aligned}
\end{equation}
Without losing any generality, we align the global reference direction of $\pmb{r}$ the same as that of $\pmb{r}_j$, e.g. both are $\pmb{e}_x$. In this case, as is exhibited in Fig.~\ref{fig:01}(c), $\Phi_{\pmb{r}-\pmb{r}_j}$ is no longer the local angular coordinate, while the real local angular coordinate measured from the local reference direction $r_0\pmb{e}_j$ should be $\Phi_{\pmb{r}-\pmb{r}_j}-\varphi_j$. To make the integral of $d\pmb{r}$ fully localized, we replace $\Phi_{\pmb{r}-\pmb{r}_j}$ by $\Phi_{\pmb{r}-\pmb{r}_j}-\varphi_j$ (correspondingly add terms outside the integral) so that Eq.~\ref{eq:H_12_1} becomes 
\begin{equation}\label{eq:H_12_2}
    \begin{aligned}
        H_{12}
        &=
        {\sum\limits_{j}}
        e^{i({\Delta}kz_j+{\Delta}m\varphi_j)}
        {\sum\limits_{j'}}
        e^{ik_2(z_{j'}-z_j)}
        e^{im_2(\varphi_{j'}-\varphi_j)}
        \\
        &{\times}
        {\int}d\pmb{r}
        A_{n_1l_1}(\pmb{r}-\pmb{r}_j)
        H(\pmb{r})
        A_{n_2l_2}(\pmb{r}-\pmb{r}_{j'})
        \\
        &{\times}
        e^{i{\Delta}m({\Phi_{\pmb{r}-\pmb{r}_j}}-\varphi_j)}
        e^{im_2[({\Phi_{\pmb{r}-\pmb{r}_{j'}}}-\varphi_{j'})-({\Phi_{\pmb{r}-\pmb{r}_j}}-\varphi_j)]},
    \end{aligned}
\end{equation}
where ${\Delta}k=k_2-k_1,{\Delta}m=m_2-m_1$. Now the integral to $d\pmb{r}$ is fully localized (also widely called the Slater-Koster coefficient), only determined by the relative site distance $\tau=j'-j$. Denoting it as $t_{12}^{\tau}$ and adopting the geometry of helical structure $z_j=a_0j,\varphi_j=j{\Delta}\varphi=j2\pi/a_0{\zeta}$ we have 
\begin{equation}\label{eq:H_12_3}
    \begin{aligned}
        H_{12}
        &=
        {\sum\limits_{j}}
        e^{i({\Delta}ka_0+{\Delta}m{\Delta}\varphi)j}
        {\sum\limits_{\tau}}
        e^{i(k_2a_0+m_2{\Delta}\varphi)\tau}
        t_{12,\tau}
        \\
        &=
        {\delta}[{\Delta}(k+mG)]
        {\sum\limits_{\tau}}
        e^{i(k+mG)a_0\tau}
        t_{12}^{\tau},
    \end{aligned}
\end{equation}
where $G={\Delta}\varphi/a_0=2\pi/a_0\zeta$ is the linear momentum quanta. As a benchmark, if we drop off the angular order of helical structure ${\Delta}\varphi=0$ so that $G=0,\varphi_j=j{\Delta}\varphi=0$, Eq.~\ref{eq:H_12_3} returns to the conventional TB hopping parameter of the non-helical system~\cite{kittel} 
\begin{equation}\label{eq:H_12_unhelical}
    \begin{aligned}
        H_{12}
        &=
        {\delta}({\Delta}k)
        {\sum\limits_{\tau}}
        e^{ika_0\tau}
        t_{12}^{\tau}.
    \end{aligned}
\end{equation}
The helical unit cell degrades into crystal unit cell, and $k$ becomes the crystal momentum. }

\textcolor{blue}{
The key difference between Eq.~\ref{eq:H_12_3} and Eq.~\ref{eq:H_12_unhelical} is the dependence on the OAM. In a non-helical crystal with crystal unit cell, the wavevector $k$ is a good quantum number, which means that no hybridization happens between two different $k$-states and the crystal momentum $\hbar{k}$ is conserved. However, in a helical system, the conserved quantity becomes the \textit{helical momentum ${\hbar}L$} defined as
\begin{equation}\label{eq:L}
    \hbar{L} = \hbar({k}+mG), \ \ \ G=2\pi/{\zeta}a_0, 
\end{equation}
which is composed of both the linear momentum and OAM, as a direct result of the screw symmetry. In the following discussion, $L$ is called the \textit{helical wavevector} since it shares the same unit with the crystal wavevector $k$. If we take the reduced crystal momentum convention $k{\in}[-G/2,G)$, the $L$-conservation becomes ${\delta}(L)={\delta}({\Delta}k){\delta}(m)$ which is the simultaneous conservation of linear momentum and OAM. The phase factor of the hopping parameter in Eq.~\ref{eq:H_12_3} is also determined by $L$ instead of only $k$, which agrees well with the previous work~\cite{DFT_TB_orbitalpolarization}, where the influence of $m$ is explained as the ``anisotropic hopping'' without a detailed discussion on its origin and the extension to general case. }

To address the physical picture of $L$-conservation, let us consider the electron transmission problem in the helical system. After an electron wave being injected into the helical material towards the $+z$-direction with the state $\psi_i=|k_i,m_i{\rangle}$, some of it will be scattered to the reflection state $\psi_s=|k_s,m_s{\rangle}$ (e.g. $k_s = -k_i$ for the back scattering). Due to the $L$-conservation, we have ${\Delta}L=(k_s-k_i)+G(m_s-m_i)={\Delta}k+G{\Delta}m=0$ so that there must be a non-zero OAM change $\hbar{\Delta}m=-\hbar{\Delta}k/G$. The output state is a superposition of both the transmission and reflection parts and could be represented as $\psi_o=\sqrt{1-r}\psi_i+\sqrt{r}\psi_s$ with the reflection coefficient $r$. A net OAM polarziation $r{\hbar}{\Delta}m=-r\hbar{\Delta}k/G$ can be found. In a semi-classical view, if the (group) velocity change between the output and input electronic wave packet is ${\Delta}v={\hbar}{\Delta}\bar{k}/m^*$ ($m^*$ is the effective electron mass), \textcolor{blue}{the corresponding OAM change becomes $r\hbar{\Delta}m=-r\hbar{\Delta}\bar{k}/G \ {\propto} \ {\zeta}{\Delta}v$}, which is proportional to the screw fold number $\zeta$ of the helical structure, reflecting its helical-induced origin.

The $L$-conservation is also possible to be vital in many-body effects. Let us consider the umklapp process as an example. In a non-helical system, an electron can only be scattered across the whole crystal Brillouin zone (BZ) if the momentum transfer is nonzero, such as in an electron-electron or electron-phonon interaction. However, in a helical system, the scattering medium can be momentum-free or in the long-wave limit, because there is another way to achieve the scattering by exchanging the linear crystal momentum and OAM. This can be done by transitioning between the states $|k,m{\rangle}$ and $|k \ {\pm} \ pG,m \ {\mp} \ p{\rangle},p{\in}\mathbb{Z}$, which does not require any external momentum supply. In terms of the Feynman diagram, this additional path can open up new possibilities for scattering processes. It can also make some previously insignificant diagrams more important. This could have a significant impact on many-body interactions, and it is worth further investigation.


\subsection{OAM inheritance}\label{sec:modeling_orbital_inheritance}

\textcolor{blue}{
Discussions on Eq.~\ref{eq:H_12_1}-\ref{eq:H_12_3} suggest that the basis like Eq.~\ref{eq:Psi_novarphi} is not suitable for helical unit cell, which is understandable because there is an extra helical symmetry need to be considered. To fix this problem, inspired by the trick we used from Eq.~\ref{eq:H_12_1} to Eq.~\ref{eq:H_12_2} about $\Phi_{\pmb{r}-\pmb{r}_j}$, we can write down the new TB basis for helical unit cell as:
\begin{equation}\label{eq:Psi_helical}
    \begin{aligned}
        \Psi_{nl,km}(\pmb{r}) 
        &= {\sum\limits_j}e^{ikz_j}A_{nl}(\pmb{r}-\pmb{r}_j)e^{im(\Phi_{\pmb{r}-\pmb{r}_j}-\varphi_j)} \\
        & = {\sum\limits_j}[e^{ikz_j}e^{-im\varphi_j}][A_{nl}(\pmb{r}-\pmb{r}_j)e^{im\Phi_{\pmb{r}-\pmb{r}_j}}] \\
        & = {\sum\limits_j}{\langle}X_j|\Omega_{km}{\rangle}\psi_{nlm}(\pmb{r}-\pmb{r}_j), 
    \end{aligned}
\end{equation} }
where ${\langle}X_j|$ is the position basis of helical unit cell $j$. In the last line, the influence of helical structure to the phase of $\Psi_{nl,km}$, or in other words the $j$-dependent part, has totally been organized into $\Omega_{km}$ defined as
\begin{equation}\label{eq:Omega}
    \Omega_{km}(z,\varphi)=e^{ikz}e^{-im\varphi}
\end{equation}
\textcolor{blue}{
and the left term $\psi_{nlm}(\pmb{r}-\pmb{r}_j)$ is the unrotated atomic orbital, the phase angle of which $\Psi_{\pmb{r}-\pmb{r}_j}$ is measured from the global reference direction $\pmb{e}_x$ without helical rotational order. It is easily checked that $\Psi_{nl,km}$ defined in Eq.~\ref{eq:Psi_helical} satisfies the Bloch theorem in helical materials under screw symmetry (as is proved in Appendix~\ref{sec:translational_symmetry}, and its hopping parameter has the same form with Eq.~\ref{eq:H_12_unhelical} (i.e. diagonal in crystal momentum). Compared with Eq.~\ref{eq:Psi_novarphi}, Eq.~\ref{eq:Psi_helical} is with the substitution of local azimuthal angle $\Phi_{\pmb{r}-\pmb{r}_j}-\varphi_j$ for the global azimuthal angle $\Phi_{\pmb{r}-\pmb{r}_j}$. This is not just a mathematical trick but reflects the physical reality that the local rotational degree of freedom, $\Phi_{\pmb{r}-\pmb{r}_j}$, is not as trivially identical for all the atomic sites as in non-helical systems, but becomes coupled to the global rotational degree of freedom $\varphi$ in helical systems. Namely the atomic OAM is \textit{``inherited''} to the helical structure. }

It is essential to emphasize that since $\Phi_{\pmb{r}-\pmb{r}_j}$ ranges continuously from 0 to $2\pi$ while $\varphi_j$ can only take discrete values $\varphi_j=2{\pi}j/\zeta, j=0,1,2,...,\zeta-1$, the OAM inheritance cannot be arbitrary because for any state with $|m'|=m+[\zeta/2]>[\zeta/2]$ (the brackets $[...]$ means rounding down), the following relation is always valid:
\begin{equation}
    \begin{aligned}
        e^{-im'\varphi_j} &= e^{{\pm}i(m+[\zeta/2])\varphi_j} = e^{{\pm}im\varphi_j}e^{{\pm}i{\pi}j} \\
        &= e^{{\pm}im\varphi_j}e^{{\mp}i{\pi}j} = e^{{\pm}i(m-[\zeta/2])\varphi_j},
    \end{aligned}
\end{equation}
which means that any state with $|m'|>[\zeta/2]$ is actually equivalent to another state with $|m'|<[\zeta/2]$. In other words, the number of distinct $m$ that can be \textit{inherited} is limited by screw symmetry order $\zeta$ as
\begin{equation}\label{eq:n_symmetry_limit}
    m=0,\pm1,\pm2,...,\pm[\zeta/2].
\end{equation}
For illustration, consider a helical system with a screw symmetry order of $\zeta=3$. In this case, only states with OAM values of $0,\pm1$ can be uniquely inherited. For example, a $d$-orbital, which typically possesses $m=\pm2$ states, would be confined to manifest as $m=\mp1$ states within this helical environment. This constraint mirrors the behavior observed in crystal momentum, where $k$ is termed quasi-momentum due to crystal lattice periodicity. Similarly, under screw symmetry, the OAM states are appropriately referred to as quasi-OAM.


\subsection{Chirality}\label{sec:modeling_chirality}

One of the important properties of helical structures is the chirality, which refers to the feature that the structure is distinguishable from its mirror image. Intuitively, the left/right-hand chiral helix can be constructed by rolling a chain in a clockwise/anticlockwise direction along the helical axis. Based on the geometric interpretation of the helical structure defined in section~\ref{sec:modeling_orbital_inheritance}, namely the linear coordinate $z_j$ and angular coordinate $\varphi_j$ of $j$-th helical unit cell, the influence of chirality to helical geometry can be reflected as the sign of the screw fold number $\zeta$, namely $\zeta>0$ and $\zeta>0$ for two different chirality. To illustrate it explicitly, we have the position coordinates of helixes with the opposite chirality as
\begin{equation}\label{eq:nu_z_varphi}
    z_j = ja_0, \ \varphi_j = \nu{j}|{\Delta}\varphi|
\end{equation}
or equivalently $z_j = -{\nu}ja_0, \ \varphi_j = {j}|{\Delta}\varphi|$, \textcolor{blue}{where $\nu=\pm1$ for different chirality respectively.}

A direct result can be found from Eq.~\ref{eq:H_12_3} that the helical momentum becomes
\begin{equation}
    \hbar{L} = \hbar(k+{\nu}m|G|).
\end{equation}
For example, $p_+$ state with $m=+1$ will be inherited to helical momentum as $\hbar(k+mG)$ and $\hbar(k-mG)$ in different chiral system respectively. However, in a system that preserves time reversal symmetry, since there are always degenerate state pairs $|k,m{\rangle}$ and $|-k,-m{\rangle}$, the helical momentum could still remain unchanged in chirality reversing.

Besides, the non-zero Berry phase arising from the inversion symmetry breaking along the helical axis is also dependent on the chirality of the helix, which can be calculated as 
\begin{equation}\label{eq:berry_gamma}
    \Gamma = {\sum\limits_{E_j<E_f}}\int_{-\pi/{\zeta}a_0}^{\pi/{\zeta}a_0}{\langle}\kappa_{j}|\frac{{\partial}}{\partial{k}}|\kappa_{j}{\rangle}dk, 
\end{equation}
where the summation is over all the occupied eigen-states $|\kappa_{j}{\rangle}$ with energy $E_j$ smaller than Fermi energy $E_f$. In the TB representation, $|\kappa_{j}{\rangle}$ is a linear combination of TB basis defined in Eq.~\ref{eq:Psi_helical} as
\begin{equation}
    |\kappa_{j}{\rangle}={\sum_{nlm}}c_{j;nl,km}|\Psi_{nl,km}{\rangle},
\end{equation}
where $c_{j;nl,km}$ are the superposition coefficients representing the hybridization between different basis and are determined by the chemical details. If the chirality is reversed from $\nu=+1$(-1) to $\nu=-1$(+1), the only change to the material structure is reversing the sign of the $z$-coordinate while the circumferential coordinate $\varphi$ is kept unchanged. This results in a change of global phase factor from $e^{ikz}$ to $e^{-ikz}$, which is equivalent to changing $k$ into $-k$. Therefore, the operator ${\partial}_k$ in Eq.~\ref{eq:berry_gamma}  becomes ${\partial}_{-k}=-{\partial}_k$. Another way to understand sign reverse is that the $z$-position operator is defined as $\hat{z}=-i{\partial}_k$. Therefore, changing the chirality from $\nu=+1$(-1) to $\nu=-1$(+1) is equivalent to changing $\hat{z}$ to $-\hat{z}$. As a conclusion, the Berry phase of structures with opposite chirality should have opposite signs. Correspondingly, the physical quantities related to the Berry phase, such as electric polarization and second-order nonlinear optical responses, e.g., shift current, will also be able to be dependent on the chirality, which makes chirality an experimentally detectable and measurable property.


\subsection{Helical potential projection}

Actually, it is straightforward to verify that the helical wavevector label a complete orthogonal basis on the helical structure. With the helical wavevector $L = k - mG$, the basis $\{|\Omega_L{\rangle}=|\Omega_{km}{\rangle}|L = k + mG, k {\in} BZ, m{\in}\mathbb{Z} \}$ satisfies 
\begin{equation}
    \begin{alignedat}{1}
        & {\sum\limits_{j}}{\langle}\Omega_L|X_j{\rangle}{\langle}X_j|\Omega_{L+{\Delta}L}{\rangle} = {\sum\limits_{j}}\Omega^*_L(z_j,\varphi_j)\Omega_{L+{\Delta}L}(z_j,\varphi_j) \\
        &= {\sum\limits_{j}}
        e^{i({\Delta}kz_j+{\Delta}m\varphi_j)} = {\sum\limits_{j}}
        e^{i{\Delta}(k+mG)a_0j} = {\delta}_{{\Delta}L},
    \end{alignedat}
\end{equation}
where $|X_j{\rangle}{\langle}X_j|$ is the projection operator from the whole cylindrical structure $\{(z,\varphi)\}$ to the subspace of the helical structure $\{(z_j,\varphi_j)|z_j=a_0j,\varphi_j=a_0Gj,j{\in}\mathbb{Z}\}$. It should be noted that ${\sum\limits}_{j}{\langle}\Omega_{L}|X_j{\rangle}{\langle}X_j|\Omega_{L'}{\rangle} {\neq} {\langle}\Omega_{L}|\Omega_{L'}{\rangle}={\delta}_{k,k'}{\delta}_{m,m'}$ since in the helical subspace we do not have ${\sum\limits}_{j}|X_j{\rangle}{\langle}X_j|=\mathbb{I}$. In this view, the coupling between $k$ and $m$ is arisen from the dimension collapse from a 2D cylindrical space into a 1D helical space.

For any potential function defined on the cylindrical space $V=V(z,\varphi)$, its value in the helix chain space is expressed as $V_j={\langle}X_j|V|X_j{\rangle}=V(z_j,\varphi_j)$. We could project it into different $L$-components as
\begin{equation}
    \begin{aligned}
        &V_j = {\sum\limits_L}V_{L}e^{iLa_0j}
    \end{aligned}
\end{equation}
with the component strength
\begin{equation}
    \begin{aligned}
        V_{L} &= {\sum\limits_j}{\langle}\Omega_{L}|X_j{\rangle}V_j = {\sum\limits_j}V_je^{-iLa_0j}.
    \end{aligned}
\end{equation}

Let $V$ become the global helical potential $V_h$, which is from the $\zeta$-fold helical structure without considering any local chemical environment, it should be a constant at all the helical unit cells as $V_{h,j}=V_{h,j{\pm}1}{\equiv}V_0$ so that 
\begin{equation}
    \begin{aligned}
        V_{h,j{\pm}1} &= {\sum\limits_L}V_Le^{iKa_0(j{\pm}1)} = {\sum\limits_L}e^{{\pm}iLa_0}V_{L}e^{iLa_0j} \\
        &= {\sum\limits_L}V_{L}e^{iLa_0j} = V_{h,j}
    \end{aligned}
\end{equation}
thus it only has the $L=0$ (modulo $2\pi$) component, namely $k=-mG$. Therefore, the helical potential $V_h$ is in the form 
\begin{equation}\label{eq:helical_potential}
    \begin{aligned}
        V_h(z,\varphi) &= {\sum\limits_m}v_{m}e^{im(Gz+\varphi)},
    \end{aligned}
\end{equation}
where the strength of different $m$-components can be obtained by projecting from the total potential function $V$ as
\begin{equation}
    \begin{aligned}
        v_m &= {\iint}V(z,\varphi)e^{-im(Gz+\varphi)}r_0dzd\varphi.
    \end{aligned}
\end{equation}
Note that here the integral is over the whole cylindrical space. This projection offers a way to quantatively obtain the strengh of different components of the helical potential from an arbitrary potential function (e.g. the Kohn-Sham potential from DFT calculation).

This result is also consistent with our discussion in Eq.~\ref{eq:H_12_3} for a helical momentum conserving. On the other hand, those components with $L{\neq}0$ are not helical momentum-conserving thus will give the single-particle eigen-states (i.e. quasi-electron) a finite life-time, which will be discussed in future works.


\subsection{Helical-induced SOC}\label{sec:modeling_SOC}

An effective electric field ${\nabla}V_h$ can be created by the helical potential regardless of the chemical details in each helical unit cell (the local chemical environment). For simplicity we only take one component of $V_h$ in Eq.~\ref{eq:helical_potential} with $m=\beta$ as $V_h=V_0\exp[i\beta(Gz+\varphi)]$. Therefore, ${\nabla}V_h$ is calculated as 
\begin{equation}
    \begin{aligned}
        {\nabla}V_h &= \pmb{e_z}\partial_z{V_h} +
        \pmb{e_t}\partial_{r_0{\varphi}}{V_h}
        =
        i\beta(G\pmb{e_z} + \frac{1}{r_0}\pmb{e_t})V_h, 
    \end{aligned}
\end{equation}
where $\pmb{e_z}$ and $\pmb{e_t}$ are unit vectors along the $z$- and circumferential directions, respectively. The SOC arising from ${\nabla}V_h$ is written as
\begin{equation}\label{eq:HSOC_general}
    \hat{H}_{SO} = -{\xi}\hat{\pmb{\sigma}}{\cdot}({\nabla}V_h{\times}\hat{\pmb{p}}) + {\rm h.c.}
\end{equation}
with a constant $\xi={g\mu_B}/{2mc}$. The momentum operator $\hat{\pmb{p}}$ can also be projected along the two directions as
\begin{equation}
    \begin{aligned}
        \hat{\pmb{p}} &= \pmb{e_z}\hat{p}_z+\pmb{e_t}\hat{p}_t \\
        \hat{p}_z &= (-i{\hbar})\partial_{z} \ \ \ 
        \hat{p}_t = (-i{\hbar})\partial_{{r_0}{\varphi}}.
    \end{aligned}
\end{equation}
We get the following expression from Eq.~\ref{eq:HSOC_general} for the SOC Hamiltonian 
\begin{equation}
    \begin{aligned}
        \hat{H}_{SO} &= -{\xi}i\beta\hat{\pmb{\sigma}}{\cdot}[(\pmb{e_z}G + \pmb{e_t}\frac{1}{r_0}){\times}(\pmb{e_z}\hat{p}_z + \pmb{e_t}\hat{p}_t)]  + {\rm h.c.}\\ 
        &= -{\xi}i\beta\hat{\pmb{\sigma}}{\cdot}\pmb{e_r}(\frac{\hat{p}_z}{r_0} - G\hat{p}_t)  + {\rm h.c.}\\ 
        &= -{\xi}i\beta\hat{\sigma}_s(\frac{\hat{p}_z}{r_0} - G\hat{p}_t) + {\rm h.c.},
    \end{aligned}
\end{equation} 
where $\pmb{e}_r=\pmb{e}_t{\times}\pmb{e}_z$ is the unit vector along the radius direction. It can be found that this helical-induced SOC (hSOC) arises from the inversion symmetry breaking along both the circumferential and helical-axis directions. The spin operator is calculated as
\begin{equation}\label{eq:sigma_spm}
    \begin{aligned}
        \hat{\sigma}_s &=\hat{\pmb{\sigma}}{\cdot}\pmb{e_r} \\
        &= \hat{\sigma}_x\cos(\varphi)+\hat{\sigma}_y\sin(\varphi) \\
        &= e^{i{\varphi}}\frac{\hat{\sigma}_x-i\hat{\sigma}_y}{2}+e^{-i{\varphi}}\frac{\hat{\sigma}_x+i\hat{\sigma}_y}{2} \\
        & = e^{i{\varphi}}{\hat{\sigma}_+}+e^{-i{\varphi}}{\hat{\sigma}}_- \\
        \hat{\sigma}_+ &= \frac{\hat{\sigma}_x-i\hat{\sigma}_y}{2} \ \ \ \
        \hat{\sigma}_- = \frac{\hat{\sigma}_x+i\hat{\sigma}_y}{2},
    \end{aligned}
\end{equation}
where $\hat{\sigma}_{x,y,z}$ are Pauli operators.

Without considering the local details (profile function), we only keep the global part of the TB basis defined in Eq.~\ref{eq:Omega} combining with the spinor $|\chi{\rangle}$ as $|\chi,k,m{\rangle}{\equiv}|\Omega_{km}{\rangle}|\chi{\rangle}=e^{ikz}e^{-im\varphi}|\chi{\rangle}$. The matrix element of $H_{SO}$ can be calculated as 
\begin{equation}\label{eq:H_so_semi}
    \begin{aligned}
        &{\langle}{\chi_1},k_1,m_1|\hat{H}_{SO}|\chi_2,k_2,m_2{\rangle} \\
        &=-{\xi}{\beta}V_0\hbar
        (\frac{k_2}{r_0}+\frac{Gm_2}{r_0})
        {\sum\limits_j}e^{i({\Delta}kz_j-{\Delta}m\varphi_j)}{\langle}\hat{\sigma}_s{\rangle} + {\rm h.c.}\\
        &=-{\xi}
        (\frac{{\beta}V_0}{r_0})
        \hbar
        L_2
        {\sum\limits_{\pm}}{\langle}\hat{\sigma}_{\pm}{\rangle}
        {\sum\limits_j}
        e^{i[{\Delta}k+({\Delta}m{\pm}1)G]a_0j}   + {\rm h.c.}\\
        &= -{\xi}
        (\frac{{\beta}V_0}{r_0})
        {\hbar}L_2
        {\sum\limits_{\pm}}
        {\langle}\hat{\sigma}_{\pm}{\rangle}
        {\delta}[{\Delta}L{\pm}G]  + {\rm h.c.}\\
        &= -{\xi}
        (\frac{{\beta}V_0}{r_0})
        {\hbar}\bar{L}
        {\sum\limits_{\pm}}
        {\langle}\hat{\sigma}_{\pm}{\rangle}
        {\delta}[{\Delta}L{\pm}G],
    \end{aligned}
\end{equation}
where ${\Delta}k=(k_2-k_1)/2,{\Delta}m=(m_2-m_1)/2,{\Delta}L=L_2-L_1,\bar{L}=L_1+L_2$. ${\langle}\hat{\sigma}_{\pm}{\rangle}={\langle}\chi_1|\hat{\sigma}_{\pm}|\chi_2{\rangle}$ suggests that $H_{SO}$ flips the spin through exchanging the spin angular momentum (SAM) with the helical momentum. The strength of $H_{SO}$ is proportional to the strength of the helical potential gradient ${\beta}V_0/r_0$, which is the effective electric field breaks the inversion symmetry in both the circumferential and the axial directions. If take the limit $r_0{\rightarrow}\infty$ the helical structure degrades into a plate thus $H_{SO}{\rightarrow}0$. It is also proportional to the helical momentum ${\hbar}\bar{L}$, just like the widely acknowledged Dresselhaus-Rashba SOC in non-helical systems~\cite{Rashba} being proportional to the crystal momentum, which again addresses that the helical momentum in helical systems plays the same role as the crystal momentum does in non-helical systems. As a conclusion, without conisdering any chemical details inside each helical unit cell, only the helical structure itself could generate a spin-flipping SOC. For the states inside the first crystal BZ $|k|<G$, ${\delta}[{\Delta}L{\pm}G]={\delta}[{\Delta}k]{\delta}[{\Delta}m{\mp}G]$, which refers to the so called orbital-mixing or channel-mixing and is able to support CISS~\cite{utsumi2020spin}.

We could further analyze the influence of hSOC in the electron transmission based on the discussion in the second last paragraph of section~\ref{sec:modeling_helical_momentum}. With the consideration of spinor $|\chi=\pm1/2{\rangle}$, the incident electronic state becomes $\psi_i=|k_i,m_i{\rangle}|{\pm}1/2{\rangle}$. According to Eq.~\ref{eq:H_so_semi}, there are two kinds of scattered states with spin-flipping: $\psi_{sp}=|k_i,m_i{\pm}1{\rangle}|{\mp}1/2{\rangle}$ and $\psi_{sc}=|k_i{\pm}G,m_i{\rangle}|{\mp}1/2{\rangle}$. As mentioned in section~\ref{sec:modeling_helical_momentum}, since the OAM-change is ${\Delta}m \ {\propto} \ {\zeta}{\Delta}v$, for a given inversion-symmetry breaking direction, e.g. 
${\Delta}v>0$, opposite chirality with different sign of $\zeta$ induces opposite OAM-change (i.e. different OAM polarization), which correspondingly leads to opposite spin-flipping through $|\psi_{sp}{\rangle}$, namely causes the chirality-dependent spin-selecticity. Therefore, CISS can be driven by either the linear momentum polarization (e.g. external electric potential bias creats ${\Delta}k$) or the OAM polarization (e.g. external magnetic field, optical vortices or just the linear momentum polarization ${\Delta}k$ as discussed in section~\ref{sec:modeling_helical_momentum}), or in one word, by the helical momentum polarization.


\subsection{TB modeling in helical systems}\label{sec:modeling_TBsummary}

In this section we discuss the advantages of helical-unit-cell scheme compared with the traditional crystal-unit-cell scheme in TB modeling. The latter scheme necessitates including TB basis functions on all constituent atomic sites within the crystal unit cell (e.g., $\zeta$ sites in the case of section~\ref{sec:modeling_orbital_inheritance}) because the TB basis (Eq.~\ref{eq:Psi_general}) only captures the translational symmetry. In contrast, the former scheme only requires considering the atoms within the smaller helical unit cell (e.g., 1 site in the case of section~\ref{sec:modeling_orbital_inheritance}) because our OAM-inherited TB basis (Eq.~\ref{eq:Psi_helical}) considers the total screw symmetry, which could significantly simplify the model, reducing the number of basis functions and facilitating deeper analysis.

However, it also should be noted that with the helical-unit-cell scheme, the $k$-space size is also $\zeta$-times larger than that of the crystal-unit-cell scheme. As a result, in principle, to comprehensively model all the electronic states, although the helical-unit-cell scheme could use less TB basis than the crystal-unit-cell one, it still needs more interaction coefficients to cope with a larger $k$-space, which is quite fair since these two schemes are on earth equivalent thus should have the same total complexity. However, since the low-energy electronic states, namely the states near CBM and VBM, are always only concentrated in a part of $k$-space, it is not necessary for a low-energy-effective model to take care of the whole $k$-space, which makes our helical-unit-cell scheme a better choice in modeling the helical systems.

We could further take use of the equivalence between the helical- and crystal-unit-cell scheme to determine the OAM of the electronic energy bands. First, we obtain the energy bands in crystal BZ with crystal unit cell, as what people commonly do in \textit{ab initio} calculations. Second, we extend the first crystal BZ into $\zeta$-times larger to obtain the helical BZ. During this process, the energy bands are also unfolded continuously to ensure a finite kinetic energy. As is discussed in section~\ref{sec:modeling_helical_momentum}, since the inherited OAM $m$ is coupled with $k$ in the helical momentum $L$, any two states $|k,m{\rangle}$ and $|k \ {\pm} \ pG,m \ {\mp} \ p{\rangle},p{\in}\mathbb{Z}$ share the same $L$ thus are identical. Therefore, the shift amount of the linear wavevector ${\Delta}k$ arisen from the band unfolding is correlated to the change of OAM quantum number as ${\Delta}m=-{\Delta}k/G$. Since the energy bands before unfolding are obtained in traditional crystal-unit-cell scheme where the OAM is not considered explicitly, they could be regarded with $m=0$ for all, so that the OAM quantum number in the helical-unit-cell scheme can be determined as $m=0+{\Delta}m=-{\Delta}k/G$. An example can be found in section \ref{subsec:OAM_InSeI}.

It's important to note that our ``simplified scenario'' with only one atomic site per helical unit cell is not an approximation. The intra-cell interactions have no OAM dependence due to the zero helical angle difference ${\Delta}\varphi=0$ in Eq.~\ref{eq:H_12_3}, thus can be simply reduced to the standard Bloch TB case. Therefore our single-site discussion in section~\ref{sec:modeling_helical_momentum} and~\ref{sec:modeling_orbital_inheritance} can be readily extended to multi-site case by taking the direct product of our single-site basis function $|\Psi_{nl,km}{\rangle}$ with the additional atomic basis functions $\{|\tau{\rangle}\}$ as $\{|\Psi_{nl,km,\tau}{\rangle}=|\Psi_{nl,km}{\rangle} |\tau{\rangle}\}$.


\section{Application to I\lowercase{n}S\lowercase{e}I}\label{sec:app in InSeI}
\subsection{OAM inheritance}\label{subsec:OAM_InSeI}

\begin{figure}
    \centering
    \includegraphics[width=0.45 \textwidth]{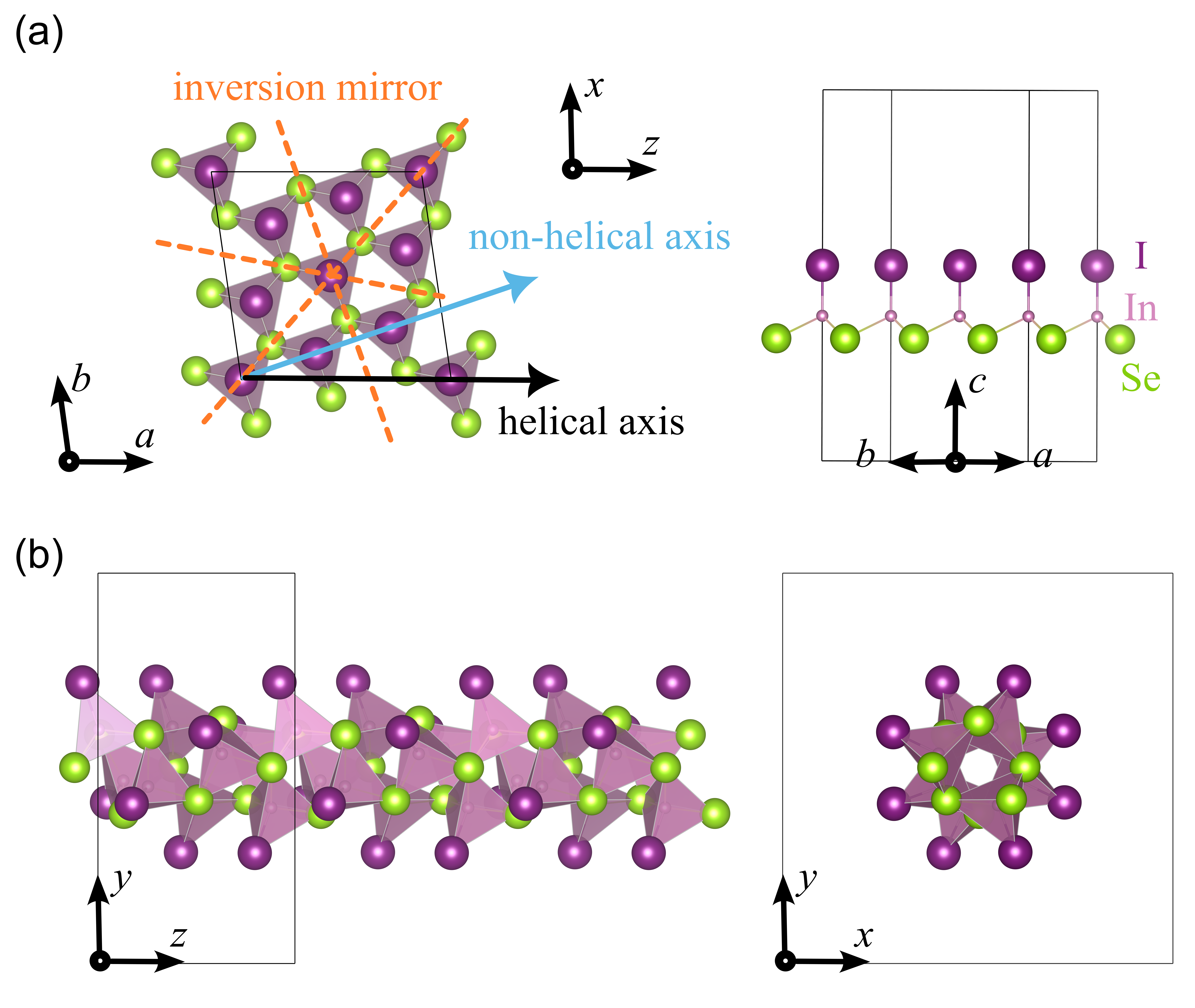}
    \caption{\label{fig:1}
    \textbf{Structure of InSeI.} (a) 2D non-helical and (b) 1D helical InSeI. }
\end{figure}

\begin{figure*}
    \centering
    \includegraphics[width=0.65 \textwidth]{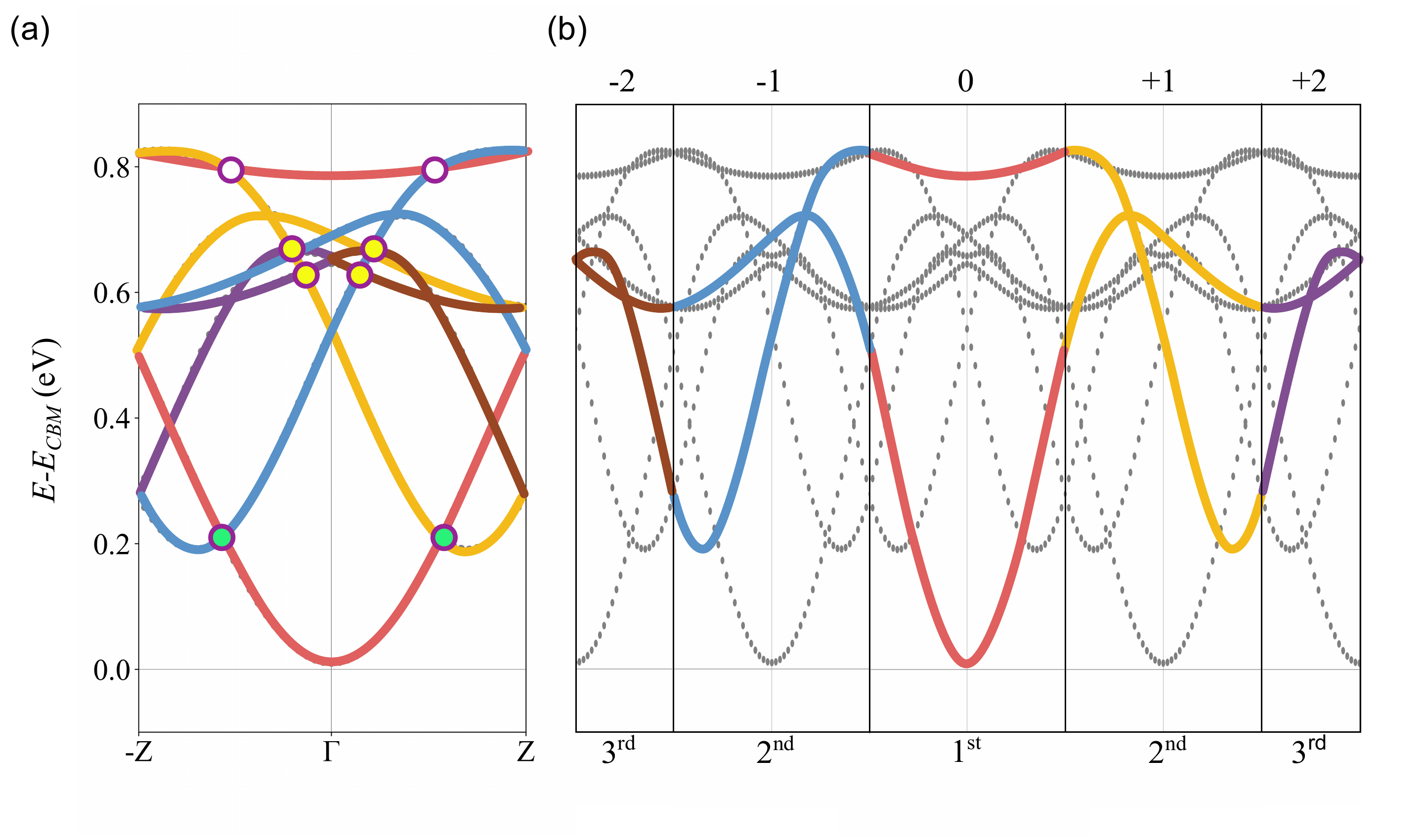}
    \caption{\label{fig:2}
    \textbf{Electronic band structure of 1D helical InSeI} in (a)crystal BZ and (b)helical BZ. 
    Different colors are used to label different OAM quantum number $m$ which is shown at the top of (b). The red, yellow and blue bands are called band-1,2 and 3 (with $m=0,+1,-1$) respectively 
    in following discussion. The yellow and green filled dots in (a) denote the degenerate points between bands with $m=\pm2,m=\pm1$ and $m=\pm1,m=\pm0$, which can be broken by hSOC. }
\end{figure*}

To clarify the structure of helical InSeI, we at first find its corresponding non-helical structure as shown in Fig.~\ref{fig:1}(a). By rolling along one of its lattice vectors (e.g. the helical axis plotted in black arrow), we can construct the 1D helical InSeI as is shown in Fig.~\ref{fig:1}(b). The 2D non-helical InSeI possesses a P3${m1}$ space group with three inversion mirrors (orange dashed lines). Since our rolling direction is not perpendicular to any of them, the inversion symmetry along the rolling axis is broken, creating a helical structure. This can also be verified by the fact that the right-handed helical InSeI has a P4$_1$ space group containing a 4-fold screw axis. 

The DFT-calculated energy bands (without SOC) of helical InSeI near the conduction band minimum (CBM) are shown in Fig.~\ref{fig:2}. We focus on the conduction band (CB) rather than the valence band (VB) because the energy of the former, i.e. the electron affinity, is close to the Fermi level of commonly used metal electrodes, while the latter is not, which makes CB more practical for device manufacturing~\cite{Zhao2023}. In Fig.~\ref{fig:2}(a), $k$ is defined within the first crystal BZ. As discussed in section \ref{sec:modeling_orbital_inheritance}, the atomic OAM can be inherited to the helical structure. To identify the OAM, we expand the crystal BZ to 4-times larger (from $2\pi/4a_0$ to $2\pi/a_0$) and unfold the bands continuously to obtain the helical BZ, which, as shown in Fig.~\ref{fig:2}(b), contains from the first to the fourth crystal BZs with labels at the panel bottom. According to the discussion in section~\ref{sec:modeling_helical_momentum} about helical momentum, the parts of helical BZ that correspond to the second, third, and fourth crystal BZs are $\pm{G}$- and $\pm2{G}$-shifted from the first crystal BZ. Therefore, the states in these parts have the corresponding helical momentum $L=\hbar (k \ {\pm} \ G)$ and $L=\hbar (k \ {\pm} \ 2G)$, with the OAM quantum number $m=\pm1$ and $m=\pm2$ respectively, as labelled on the panel top of Fig.~\ref{fig:2}(b). It can be found that the sign of $m$ is related to that of $k$, so that $L(k)=\hbar(k+mG)=-L(-k)$, which satisfies the time-reversal symmetry as is discussed in section~\ref{sec:modeling_SOC}. 

\begin{figure}
    \centering
    \includegraphics[width=0.4 \textwidth]{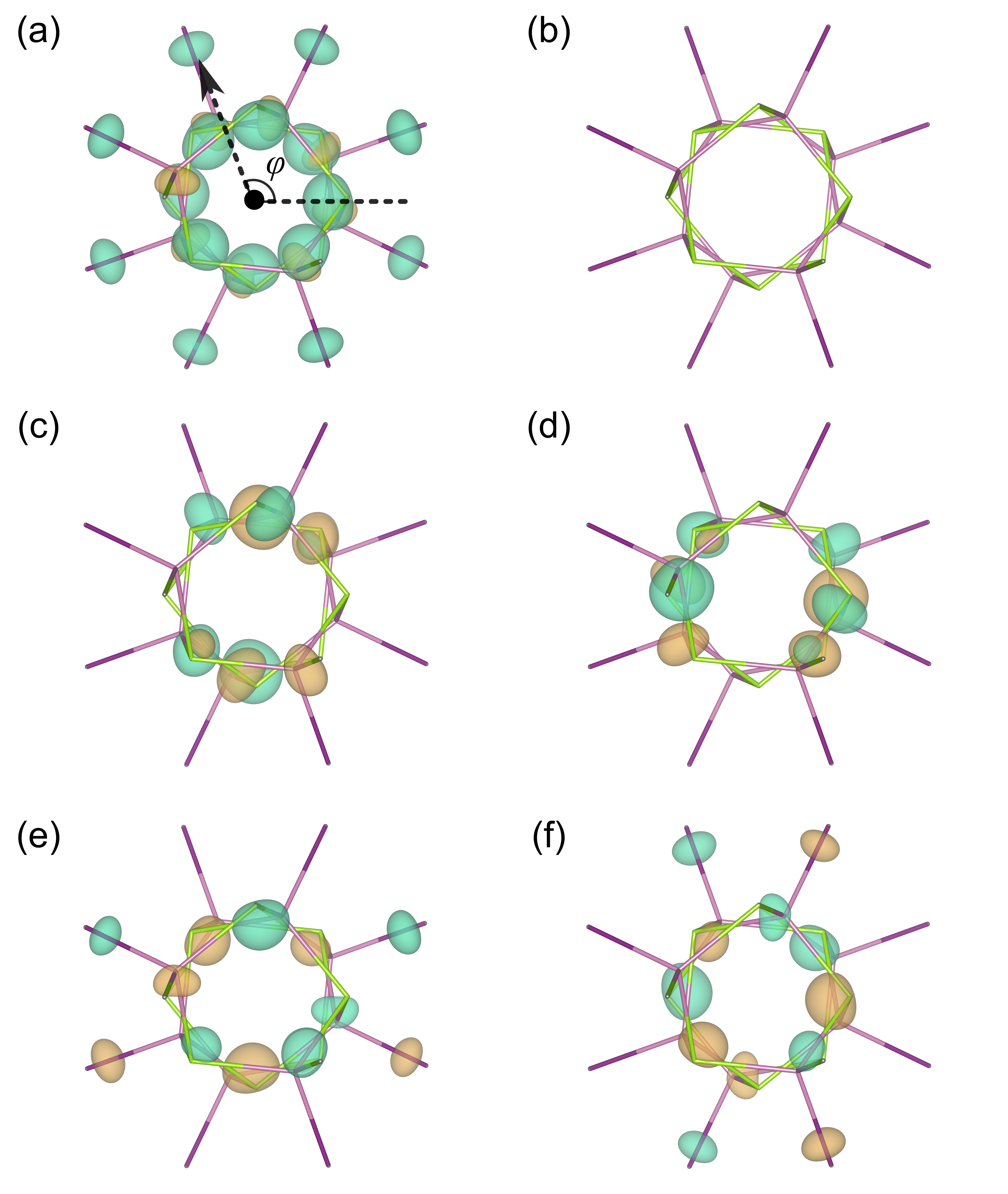}
    \caption{\label{fig:3}\textbf{OAM inheritance observed in DFT-calculated wavefunctions.} \textcolor{blue}{The isosurface value is +0.12 in yellow and -0.12 in green (unit:\AA$^{-\frac{3}{2}}$), namely the two colors represent opposite phases. (a) Real and (b) imaginary part of Band-1, with no phase dependence on $\varphi$ (i.e. $m=0$ for $e^{im\varphi}$). (c) Real and (d) imaginary part of band-2, with non-zero phase dependence on $\varphi$ (i.e. $m{\neq}0$ for $e^{im\varphi}$). (e) Real and (f) imaginary part of band-1, but in the second crystal BZ, also with non-zero phase dependence on $\varphi$. Only one $k$ point is arbitrarily selected for each band but the key features are not dependent on $k$. } }
\end{figure}

In a helical system, it is not appropriate to decompose the electronic states into globally identical spherical harmonics ($s,p_x,p_y...$) as is commonly done to obtain orbital-projected energy bands in \textit{ab initio} calculations, which is because of the tilted local chemical environment as discussed in section~\ref{sec:modeling_TB_basis}. Instead, we can directly observe the phase evolution of the wavefunction along the circumferential direction to verify the OAM. As shown in Fig.~\ref{fig:3}(a)(b), the wavefunction of band-1 has neither a significant imaginary component nor a strong sign dependence (i.e. phase dependence) on $\varphi$, which corresponds to a zero OAM $m=0$ so that the phase factor $e^{i0\varphi}{\equiv}1$. In contrast, as shown in Fig.~\ref{fig:3}(c)(d), the real and imaginary components of band-2/3 are almost identical, and both have significant phase variation depending on $\varphi$. The only difference is their pointing directions that orthogonal to each other. By comparing with atomic orbital $p_x, p_y$ in Eq.~\ref{eq:p_xyz_orbitals}, which are the real and imaginary component of $p_{\pm}$ orbital (with $m=\pm1$) respectively, we can verify that band-2/3 has a typical $|m|=1$ OAM with the phase factor $e^{{\pm}i\varphi}$. Besides, according to the definition of helical momentum defined in Eq.~\ref{eq:L} and the discussion in last paragraph, band-1 in the second crystal BZ should have an OAM of $m=\pm1$, which can also be observed in Fig.~\ref{fig:3}(e)(f). These results strongly support the validity of our general discussion in section~\ref{sec:modeling_orbital_inheritance} and~\ref{sec:modeling_helical_momentum}.

\subsection{Helical-induced SOC}\label{subsec:hSOC_InSeI}

\begin{figure*}
    \centering
    \includegraphics[width=0.7 \textwidth]{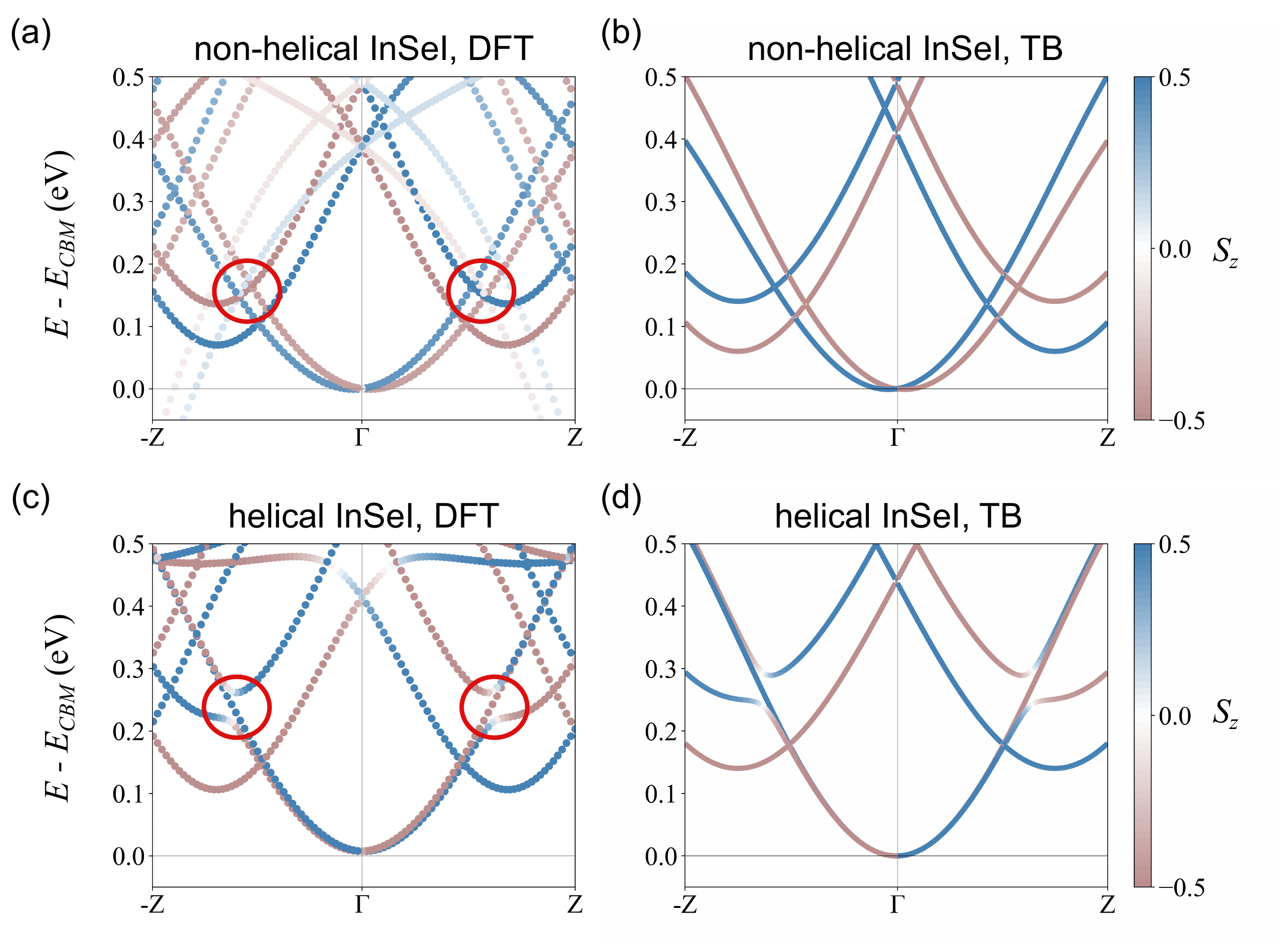}
    \caption{\label{fig:4}
    \textbf{Helical-induced SOC in InSeI.}
    (a) DFT and (b) TB-model band of non-helical InSeI along the $\pmb{b}{\times}\pmb{c}$ direction, where $\pmb{b},\pmb{c}$ are lattice vectors defined in Fig.~\ref{fig:1}(a), and $\pmb{b}{\times}\pmb{c}$ is approximately along the helical axis shown in Fig.~\ref{fig:1}(a). (c) DFT and (d) TB-model band of helical InSeI. The color bar represents the expectation value of $z$-direction spin $S_z$. \textcolor{blue}{The red circles highlight the spin-mixing splitting due to the hSOC.} }
\end{figure*}

With the consideration of SOC, the DFT-calculated band structures of non-helical and helical InSeI are shown in Fig.~\ref{fig:4}(a) and (c), respectively. Both systems have a large spin-conserving spin splitting between band-2 and band-3, which is a typical result of atomic SOC between $p_x,p_y$ orbits~\cite{Zhao2023}. However, in helical InSeI, there are spin-mixing energy gaps highlighted by red circles in Fig.~\ref{fig:4}(c), where the spin order becomes discontinuous, suggesting that there is an additional spin-flipping SOC that only exists in helical InSeI instead of non-helical InSeI. 

We can verify that this spin-flipping SOC is able to be well explained by our discussion about hSOC in section~\ref{sec:modeling_SOC} from three aspects. First, according to Eq.~\ref{eq:H_so_semi}, hSOC can only happen between states with ${\Delta}m=\pm1$. As shown in Fig.~\ref{fig:2}(a), there are four degenerate points between $m=\pm1$, $m=\pm2$ bands (highlighted in yellow filled dots) and four degenerate points between $m=\pm1$,$m=0$ bands (highlighted in green and white filled dots) that satisfy this restriction. The two green-filled ones, which are located in the low-energy region that we are interested, are broken into spin-mixing energy gaps by hSOC as highlighted by red circles in Fig.~\ref{fig:4}(c), which agrees well with our discussion about helical momentum conservation and hSOC. Second, still take these spin-mixing energy gaps as an index, after exerting a 6\% tensile strain  to helical InSeI along the helical axis, as is shown in Fig.~S1(a), the gap size is significantly reduced. It could be observed that the tensile strain descends the energy of band-2/3 relative to band-1, which shifts the two spin-mixing gaps towards the $\Gamma$ and reduces their $|k|$ value. According to Eq.~\ref{eq:H_so_semi}, $H_{SO}$ is proportional to $|k|$, so does the size of the hSOC-related spin-mixing gap. Therefore, this strain regulation to spin-mixing gaps of InSeI could be well explained by our analysis.

Finally, as a more detailed verification, we build a low-energy-effective TB model for band-1,2,3 based on the discussion in section~\ref{sec:modeling_orbital_inheritance} and \ref{sec:modeling_helical_momentum}. There are three basis states: two $p_x,p_y$-type ($|m|=1$) denoted as $|x{\rangle},|y{\rangle}$, and one $s$ or $p_z$-type ($|m|=0$) denoted as $|z{\rangle}$, and each with two spin components, thus totally six orbits in this model (details can be found in section~\ref{sec:TB details}). Roughly fitting our model to the DFT band structure, we can see in Fig.~\ref{fig:4}(b)(d) that our model is able to give successful interpretation to the dispersion, parity, spin texture, and spin splitting. Therefore, it can be concluded that our analysis about helical-induced SOC works well in explaining DFT results, which proves the validity of our corresponding discussion in section~\ref{sec:general}.

\begin{figure*}
    \centering
    \includegraphics[width=0.7\textwidth]{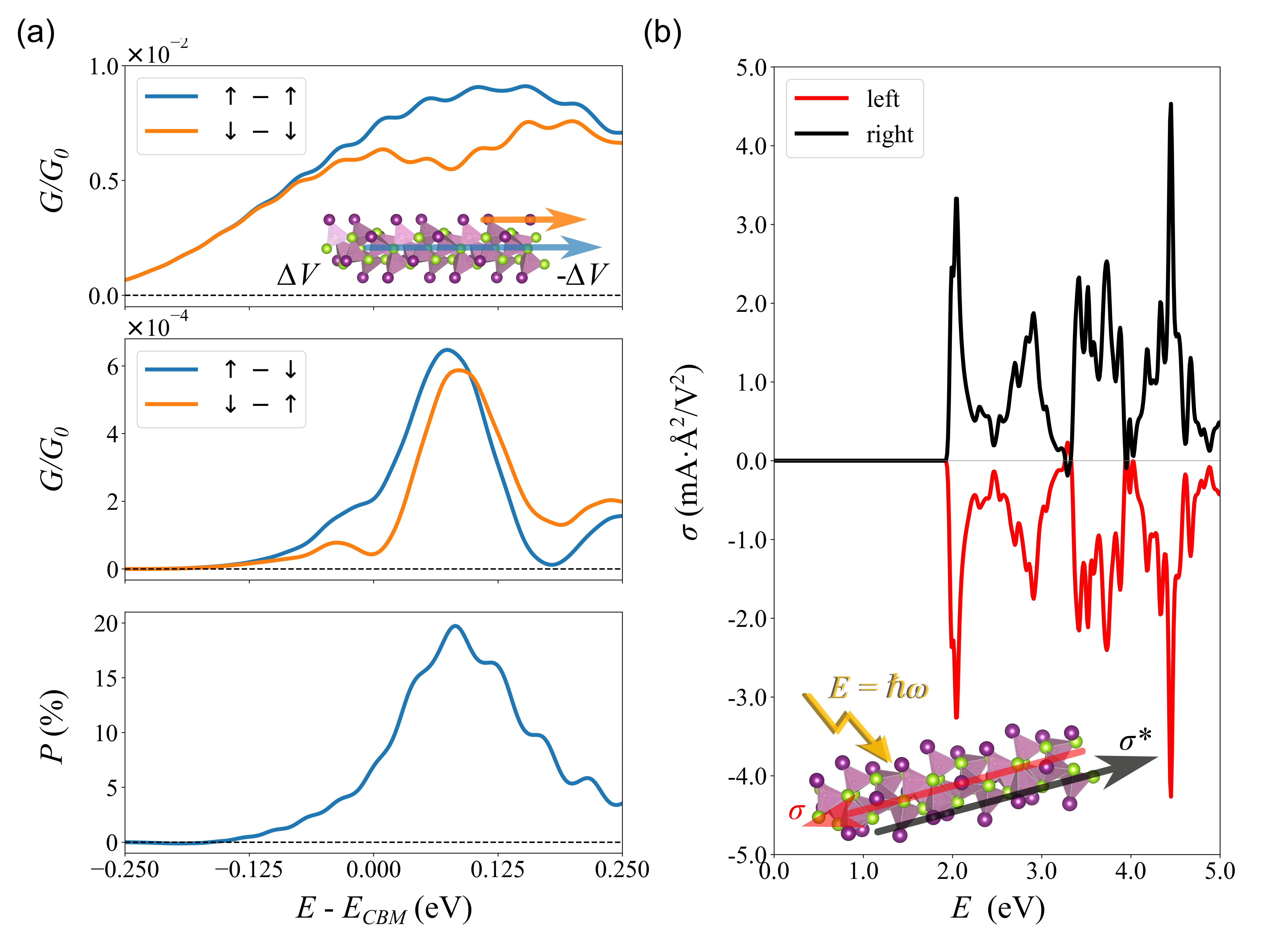}
    \caption{\label{fig:6}\textbf{Helical-induced polarization of helical InSeI.}
    (a) Ballistic conductance and spin polarization arising from hSOC in helical InSeI (with 6\% tensile strain as is shown in Fig.~S1). The spin-conserving and spin-flipping conductances are polarized as  $G_{\uparrow{\uparrow}} > G_{\downarrow{\downarrow}}$ and $G_{\uparrow{\downarrow}} {\neq} G_{\downarrow{\uparrow}}$, a non-zero spin polarization $P$ can be achieved. If the hSOC turned off, $G_{\uparrow{\uparrow}} = G_{\downarrow{\downarrow}}$ and $G_{\uparrow{\downarrow}}= G_{\downarrow{\uparrow}}=0$, leading to a zero polarization. (b) Chirality dependent shift current conductivity $\sigma^{zzz}$ induced by non-zero Berry connection in helical InSeI.}
\end{figure*}

\subsection{Ballistic spin-polarized electron transport}\label{subsec:helical-CISS}

Helical InSeI is a promising material for spintronic devices based on the spin-flipping hSOC. Consider an InSeI nanowire with a length of 80 unit cells. If we connect this nanowire to two electrodes and apply a small bias voltage ${\Delta}V$, we can create an electric current flowing through the nanowire. According to the Landauer-Buttiker formalism for ballistic transport~\cite{ballistic_book}, the electric current can be calculated as
\begin{equation}
    \begin{aligned}
        I &= \frac{1}{e}\int_{E_f-{\Delta}V}^{E_f+{\Delta}V}G(E)[f_1(E)-f_2(E)]dE \\
        &{\approx} \ G(E_f){\Delta}V/e
    \end{aligned} 
\end{equation}
where $G(E)$ is the conductance contributed by the electronic states with energy $E$ and $f_{1,2}(E)$ is the Fermi-Dirac distribution function in electrode 1,2. The approximate equality is valid for small ${\Delta}V$, so that the current is simply proportional to $G(E_f){\equiv}G$, which means that only the states near $E_f$ contribute to conduction. The conductance for spin-up/down electrons moving from one electrode to another without changing their spin states is denoted by $G_{{\uparrow}{\uparrow}}$/$G_{{\downarrow}{\downarrow}}$, which is called the spin-conserved conductance. Similarly, the spin-flipping conductance is denoted by $G_{{\downarrow}{\uparrow}},G_{{\uparrow}{\downarrow}}$. To characterize the spin selectivity, we also define the spin polarization of electronic current as
\begin{equation}
    \begin{aligned}
        P&=\frac{I_{{\uparrow}{\uparrow}}+I_{{\downarrow}{\uparrow}}-I_{{\downarrow}{\downarrow}}-I_{{\uparrow}{\downarrow}}}{I_{{\uparrow}{\uparrow}}+I_{{\downarrow}{\uparrow}}+I_{{\downarrow}{\downarrow}}+I_{{\uparrow}{\downarrow}}} \\
        & {\sim}\frac{G_{{\uparrow}{\uparrow}}+G_{{\downarrow}{\uparrow}}-G_{{\downarrow}{\downarrow}}-G_{{\uparrow}{\downarrow}}}{G_{{\uparrow}{\uparrow}}+G_{{\downarrow}{\uparrow}}+G_{{\downarrow}{\downarrow}}+G_{{\uparrow}{\downarrow}}}
    \end{aligned}
\end{equation}
If $P>0$($P<0$) the spin-up(down) current is larger, or in other words, the spin-up(down) conductance is larger, so that the current carries net spin. Otherwise, the current is spin-degenerate. 

Based on the TB model built in Appendix~\ref{sec:TB details}, we calculate the $G$ and $P$ for InSeI under 6\% tensile strain (in order to lower the energy of hSOC-activated band-2/3). As shown in Fig.~\ref{fig:6}(a), both the spin-conserved and spin-flipping conductance are polarized, which generates a spin-polarized current with non-zero $P$. The polarization peak is achieved with $E_f-E_{CBM}$ around 0.1-0.125~eV, \textcolor{blue}{where $E_f$ is near the band edge of the hSOC-activated band-2/3, as is exhibited in Fig.~S1.} All these spin polarizations vanish if we turn off the hSOC, even though the intrinsic atomic SOC is still turned on and the spin-conserving energy gap between band-2/3 still exists. Therefore, it is the hSOC that activates the spin-splitting band in helical InSeI and creates the spin-polarized response to the external electrostatic field. Additionally, since the strength of hSOC is irrelevant to the chirality as shown in Eq.~\ref{eq:H_so_semi}, the spin-selectivity will be reversed for opposite chirality because changing chirality will reverse the $z$-direction spin from spin-up/down to spin-down/up. To be more specific, if the left-handed helical InSeI has a higher spin-up conductivity, then the right-handed InSeI should have a higher spin-down conductivity, which can be used to distinguish chirality of helical InSeI or degisn spintronic devices with different spin-selectivity. 

\subsection{Helical-induced shift current}\label{subsec:helical-SC}

A non-zero Berry phase in helical InSeI can be characterized by the DFT-calculated polarization $\pm3.098$ e{\AA} with opposite signs for different chiralities. This results in a shift current under a linearly polarized monochromatic electric field $\boldsymbol{E}(t)=\boldsymbol{E}(\omega) e^{-i \omega t}+c.c.$, denoted by $j_{\operatorname{sc}}^a=2 \sigma^{a b b}(0; \omega,-\omega) \operatorname{Re}\left[E_b(\omega) E_b(-\omega)\right]$, which is a second-order nonlinear photocurrent response that flows along the helical axis in opposite directions for different chiralities. The conductivity is given by~\cite{sipe2000second, wang2022generalized}
\begin{equation}\label{eq:sc}
\begin{aligned}
\sigma^{a b b}(0 ; \omega,-\omega) & =-\frac{ \pi e^3}{\hbar^2} \int \frac{d\boldsymbol{k}}{(2\pi)^d} \sum_{nm} f_{n m}R_{mn}^{a,b}r_{nm}^b r_{nm}^b \\
                                & \times \delta\left(\omega_{nm}-\omega\right).
\end{aligned}
\end{equation}
Here, $f_{n m}=f_n-f_m$ and $\hbar \omega_{n m}=\varepsilon_n -\varepsilon_m$ denote the difference in occupation numbers and energies between bands $n$ and $m$, respectively. $d$ is the dimension. $r_{nm}^b=i\left\langle u_{\boldsymbol{k} n} \mid \partial_b u_{\boldsymbol{k} m}\right\rangle$ and $\mathcal{A}_n^b=i\left\langle u_{\boldsymbol{k} n} \mid \partial_b u_{\boldsymbol{k} n}\right\rangle$ are the interband and intraband Berry connections. $R_{m n}^{a, b}(\boldsymbol{k})=-\partial_{a} \arg r_{m n}^b(\boldsymbol{k})+\mathcal{A}_m^a(\boldsymbol{k})-\mathcal{A}_n^a(\boldsymbol{k})$ is the gauge invariant shift vector. It is clear from this expression that the shift vector is proportional to the difference in Berry connections. Therefore, reversing the chirality in helical InSeI will reverse the shift current along the helical axis.

To calculate the conductivity $\sigma$, we first constructed quasiatomic Wannier functions and a first-principles TB Hamiltonian from Kohn–Sham wavefunctions and eigenvalues, using the maximal similarity measure with respect to pseudoatomic orbitals. This yielded 192 quasiatomic Wannier functions, which consist of combinations of $s$ and $p$ orbitals derived from In, Se, and I atoms. We then used the developed TB Hamiltonian to compute the nonlinear shift current, using a dense k-point sampling of 1 $\times$ 1 $\times$ 100. A small imaginary smearing factor of $\eta$ = 0.025 eV was used for the Dirac delta function integration. The calculated $\sigma$ is shown in Fig.~\ref{fig:6}(b), where a clear sign change for different chiralities is observed. This result not only agrees with our discussion in section~\ref{sec:modeling_chirality} but also proves that the shift current is an effective way to detect chirality.


\section{Conclusions}\label{sec:conclusion}
In this paper, we investigated the electronic states in helical systems and verified our findings using DFT calculations. Our investigation has revealed several distinctive properties inherent to helical systems. Firstly, helical structures can inherit the atomic OAM of their constituent atoms. Secondly, the conserved quantity in helical systems with screw symmetry is the helical momentum, which is composed of both crystal momentum and OAM. Thirdly, the helically arranged atoms create a helical potential that induces spin-flipping SOC. The inherent chirality of the helical structure enables a chiral-switchable Berry phase, which leads to opposite shift current responses in helical InSeI with different chiralities. These properties suggest that helical systems have rich potential for applications in spintronics, optoelectronics, and beyond. Our work is expected to shed light on further exploration of the distinctive properties and versatile applications of helical systems.

\begin{acknowledgments}
    The work of J.H, S.Z., and W.L. is supported by NSFC under Grant No. 62374136. H.W. thanks the support from NSFC with Grant No. 12304049. W.L. also acknowledges the support by Research Center for Industries of the Future at Westlake University under Award No. WU2022C041. The authors thank the High-Performance Computing Center of Westlake University for technical assistance.
\end{acknowledgments}

\appendix

\textcolor{blue}{
\section{Translational symmetry}\label{sec:translational_symmetry}
If act the screw operation to the TB basis in the helical unit cell $\Psi_{nl,km}(\pmb{r})$ defined in Eq.~\ref{eq:Psi_helical}, the position vector $\pmb{r}$ and angular coordinate $\Psi_{\vec{r}-\vec{r}_j}$ are transformed as $\pmb{r}{\rightarrow}\pmb{r}+a_0\pmb{e}_z+r_0{\Delta}\pmb{e}_j$ and $\Phi_{\pmb{r}-\pmb{r}_j}{\rightarrow}\Phi_{\pmb{r}-\pmb{r}_{j-1}}+{\Delta}\varphi$, so that we have
\begin{equation}
    \begin{alignedat}{1}
        &S_z{\cdot}\Psi_{nl,km}(\pmb{r}) \\
        &= S_z{\cdot}{\sum\limits}_je^{ikz_j}A_{nl}(\pmb{r}-\pmb{r}_{j})e^{im(\Phi_{\pmb{r}-\pmb{r}_j}-\varphi_j)} \\
        &= {\sum\limits}_je^{ikz_j}A_{nl}(\pmb{r}+a_0\pmb{e}_z+r_0{\Delta}\pmb{e}_{j}
        -z_j\pmb{e}_z-r_0\pmb{e}_{j})
        \\
        &\ \ \ {\times}
        e^{im(\Phi_{\pmb{r}-\pmb{r}_{j-1}}+{\Delta}\varphi-\varphi_j)}.
    \end{alignedat}
\end{equation}
With the relation $z_{j-1}\pmb{e}_z=(z_j-a_0)\pmb{e}_z$, $\varphi_{j-1}=\varphi_j-{\Delta}\varphi$ and $r_0\pmb{e}_{j-1}=r_0(\pmb{e}_{j}-{\Delta}\pmb{e}_{j})$, we furthre have 
\begin{equation}
    \begin{alignedat}{1}
        &S_z{\cdot}\Psi_{nl,km}(\pmb{r}) \\
        &= {\sum\limits}_je^{ikz_j}A_{nl}(\pmb{r}-z_{j-1}\pmb{e}_z-r_0\pmb{e}_{j-1})e^{im(\Phi_{\pmb{r}-\pmb{r}_{j-1}}-\varphi_{j-1})} \\
        &= e^{ika_0}{\sum\limits}_je^{ikz_{j-1}}A_{nl}(\pmb{r}-\pmb{r}_{j-1})e^{im(\Phi_{\pmb{r}-\pmb{r}_{j-1}}-\varphi_{j-1})} \\
        &= e^{ika_0}\Psi_{nl,km}(\pmb{r}),
    \end{alignedat}
\end{equation}
which shows that the $\Psi_{nl,km}$ satisfies the screw symmetry just as the common Bloch TB basis satisfies the translational symmetry. }

Considering the macroscopic periodic condition (Born-von Karman boundary condition), the wavefunction should be invariant after being translated through a distance of $N{\zeta}a_0$ ($N$ is a large integer number) 
\begin{equation}\label{eq:def_BZ}
    \begin{aligned}
        &\Psi_{nl,km}(\pmb{r}+N{\zeta}a_0\pmb{e}_z) \\
        &= T[N{\zeta}a_0]{\cdot}\Psi_{nl,km}(\pmb{r}) \\ 
        &= S_z[N{\zeta}a_0, N{\zeta}{\Delta}\varphi]{\cdot}\Psi_{nl,km}(\pmb{r}) \\ 
        &= S_z[a_0, {\Delta}\varphi]^{N{\zeta}}{\cdot}\Psi_{nl,km}(\pmb{r}) \\ 
        &= e^{ikN{\zeta}a_0}\Psi_{nl,km}(\pmb{r})\\
        &= \Psi_{nl,km}(\pmb{r}),
    \end{aligned}
\end{equation}
which gives us the $1^{\rm st}$ BZ of crystal unit cell
\begin{equation}
    e^{ikN{\zeta}a_0} = 1 \ \ {\Rightarrow} \ \ k{\in}[-G/2,G/2].
\end{equation}


\section{Details of TB modeling}\label{sec:TB details}

As is defined in section~\ref{subsec:hSOC_InSeI}, $|x{\rangle}$ and $|y{\rangle}$ are $p_x,p_y$-type with the orbital angular momentum component $m=\pm1$. 
\begin{equation}
    \begin{aligned}
        |x{\rangle} &= \frac{1}{\sqrt{2}}(|-1{\rangle}-|1{\rangle}) \\
        |y{\rangle} &= \frac{i}{\sqrt{2}}(|-1{\rangle}+|1{\rangle}).
    \end{aligned}
\end{equation}

InSeI has a 4-fold screw symmetry ($\zeta=4$), so that the directional polarization difference between two adjenct helical sites as is discussed in section~\ref{sec:modeling_orbital_inheritance} becomes ${\Delta}\varphi=2\pi/4=\pi/2$. According to Eq.~\ref{eq:H_12_3}, we have
\begin{equation}
    \begin{alignedat}{1}
        &{\langle}+1|H|+1{\rangle} \\
        &= \frac{t_1}{2}(e^{i\pi/2}e^{ika}+e^{-i\pi/2}e^{-ika}) \\
        &= \frac{t_1}{2}(ie^{ika}-ie^{-ika}) \\
        & = t_1\sin(ka) \\
        & = -{\langle}-1|H|-1{\rangle},
    \end{alignedat}
\end{equation}
where $t_1$ is the Slater-Koster coefficient. Within the first BZ, hopping is forbidden for ${\Delta}m\neq0$ so that
\begin{equation}
    {\langle}\pm1|H|\mp1{\rangle} = 0.
\end{equation}
Therefore the hopping paramter between $|x{\rangle}$ and $|y{\rangle}$ can be calculated as
\begin{equation}\label{eq:h_xy}
    \begin{aligned}
        h_{xy} &= {\langle}x|H|y{\rangle} \\
        &= \frac{i}{2}({\langle}-1|-{\langle}+1|)H(|-1{\rangle}+|+1{\rangle}) \\
        &= i\frac{{\langle}-1|H|-1{\rangle}-{\langle}+1|H|+1{\rangle}}{2} \\
        &= it_1\sin(ka),
    \end{aligned}
\end{equation}

\begin{equation}
    \begin{aligned}\label{eq:h_xx}
        h_{xx} &= {\langle}x|H|x{\rangle} \\
        &= \frac{1}{2}({\langle}-1|-{\langle}+1|)H(|-1{\rangle}-|+1{\rangle}) \\
        &= \frac{{\langle}-1|H|-1{\rangle}+{\langle}+1|H|+1{\rangle}}{2} \\
        &= 0 = h_{yy}.
    \end{aligned}
\end{equation}
Recall that $|x{\rangle}$ and $|y{\rangle}$ are directional polarized to global $x$- and $y$-direction respectively. Therefore the TB Hamiltonian of $(|x{\rangle},|y{\rangle})^{\pmb{T}}$ subspace ${H_{xy}}$ should satisfy the 4-fold rotational symmetry, or in other words, be invariant under 90${^{\circ}}$-rotational operation along ${z}$-axis as
\begin{equation}
    \begin{aligned}
    H_{xy} &=
    \begin{bmatrix} 
        h_{xx} & h_{xy} \\
        h_{xy}^* & h_{yy}
    \end{bmatrix} \\
    &= R^{-1}H_{xy}R \\
    &= \begin{bmatrix} 
        0 & -1 \\
        1 & 0
    \end{bmatrix}
    \begin{bmatrix} 
        h_{xx} & h_{xy} \\
        h_{xy}^* & h_{yy} 
    \end{bmatrix}
    \begin{bmatrix} 
        0 & 1 \\
        -1 & 0 
    \end{bmatrix} \\
    &= \begin{bmatrix} 
        h_{yy} & -h_{xy}^* \\
        -h_{xy} & h_{xx} 
    \end{bmatrix} \\ 
    &{\Rightarrow} h_{xx}=h_{yy}, \ \ \text{Re}[V_{xy}]=0,
    \end{aligned}
\end{equation}
and Eq.~\ref{eq:h_xy}, \ref{eq:h_xx} agrees well with this requirement. Since $|z{\rangle}$ has no OAM $m=0$ thus its self hopping is simply
\begin{equation}
    h_{zz} = {\langle}z|H|z{\rangle} = \frac{t_2}{2}(e^{ika}+e^{-ika}) = t_2\cos(ka),
\end{equation}
and also has no hopping with $|x{\rangle},|y{\rangle}$ due to the non-zero ${\Delta}m$.

According to Eq.~\ref{eq:H_so_semi}, the angular momentum conservation requires that hSOC is only non-zero for $|{\Delta}m|=1$, thus ${\langle}\pm1|\hat{H}_{SO}|\pm1{\rangle}=0$ so that ${\langle}x|\hat{H}_{SO}|y{\rangle}={\langle}y|\hat{H}_{SO}|x{\rangle}=0$. The only non-zero term is between $|x,y{\rangle}$ and $|z{\rangle}$ as 
\begin{equation}
    \begin{aligned}
    {\langle}x|\hat{H}_{SO}|z{\rangle}
    &= \frac{1}{\sqrt{2}}({\langle}-1|\hat{H}_{SO}|0{\rangle}-{\langle}+1|\hat{H}_{SO}|0{\rangle}) \\
    &= \lambda_R({\sigma}_{+}-{\sigma}_{-}) \\
    &= i\lambda_R{\sigma_y},
    \end{aligned}
\end{equation}

\begin{equation}
    \begin{aligned}
    {\langle}y|\hat{H}_{SO}|z{\rangle}
    &= \frac{i}{\sqrt{2}}({\langle}-1|\hat{H}_{SO}|0{\rangle}+{\langle}+1|\hat{H}_{SO}|0{\rangle}) \\
    &= i\lambda_R({\sigma}_{+}+{\sigma}_{-}) \\
    &= i\lambda_R{\sigma_x},
    \end{aligned}
\end{equation}
where the relation between $\sigma_{x,y}$ and $\sigma_{\pm}$ is obtained from Eq.~\ref{eq:sigma_spm}. The strength of hSOC is denoted as $\lambda_R$. Besides, there is also an intrinsic atomic SOC between $|x{\rangle},|y{\rangle}$ with the form $\lambda_0\sigma_z$ which is irrelevant to the helical structure~\cite{SOC_matrix}. 

Based on all the above analysis, the TB model with the basis 
\[(|x_{\uparrow}{\rangle},|y_{\uparrow}{\rangle},|z_{\uparrow}{\rangle},|x_{\downarrow}{\rangle},|y_{\downarrow}{\rangle},|z_{\downarrow}{\rangle})^{\textbf{T}} \]
is finally obtained as
\begin{equation}\label{eq:H_summary}
    H = 
    \begin{bmatrix}
        H_{{\uparrow}{\uparrow}} & H_{{\uparrow}{\downarrow}} \\
        H_{{\downarrow}{\uparrow}} & H_{{\downarrow}{\downarrow}}
    \end{bmatrix}
\end{equation}

\begin{equation}\label{eq:H_uu}
    H_{{\uparrow}{\uparrow}} = 
    \begin{bmatrix}
        U_{1} & it_{1}\sin(ka)+i{\lambda_0} & 0 \\
        {\dagger} & U_{1} & 0 \\
        {\dagger} & {\dagger} & U_{2}+t_{2}\cos(ka)
    \end{bmatrix}
\end{equation}

\begin{equation}\label{eq:H_dd}
    H_{{\downarrow}{\downarrow}} = 
    \begin{bmatrix}
        U_{1} & it_{1}\sin(ka)-i{\lambda_0} & 0 \\
        {\dagger} & U_{1} & 0 \\
        {\dagger} & {\dagger} & U_{2}+t_{2}\cos(ka)
    \end{bmatrix}
\end{equation}

\begin{equation}\label{eq:H_ud}
    H_{{\uparrow}{\downarrow}} = 
    \begin{bmatrix}
        0 & 0 & {\lambda_R} \\
        0 & 0 & i{\lambda_R} \\
        -{\lambda_R} & -i{\lambda_R} & 0
    \end{bmatrix}
\end{equation}

\begin{equation}\label{eq:H_du}
    H_{{\uparrow}{\downarrow}} = H_{{\downarrow}{\uparrow}}^{\dagger},
\end{equation}
where ${\lambda_R}={\lambda_{R0}}ka$. Roughly fitting TB band to DFT results, we have $\lambda_0 = -0.06 , \lambda_{R0} = 0.45\lambda_0,U_1 = 0.5, U_2 = U_1 - 0.15$. For equilibrium InSeI $t_1 = -0.3, t_2 = -0.35$ and for 6\% tensile strain InSeI $t_1 = -0.425, t_2 = -0.25$ (with unit eV).

\section{Quantum ballistic transport}\label{sec:ballistic details}

Through retarded Green's function $G^{R}$ and transmission matrix $\bar{T}$ we could obtain ballistic conductance $G(E)$ 
\begin{align*}
    & G(E) = G_0\bar{T}(E) \\
    & \bar{T}(E) = {\rm Tr}[{\Gamma_1}G^R{\Gamma_2}G^{R,\dagger}] \\
    & {\Gamma_{1,2}} = i[{\Sigma_{1,2}}-{\Sigma_{1,2}^{\dagger}}] = -2{\rm Im}\{\Sigma_{1,2} \} \\
    & G^R = [E-H_C-{\Sigma_{1}}-{\Sigma_{2}}]^{-1},
\end{align*}
where $G_0 = 2e^2/\hbar$ is the conductance quantum and $\Sigma_{1,2}$ is the self-energy of electrode 1,2. Approximately if only consider one conducting state with two spin components for each electrode we obtain
\begin{equation}
    {\Sigma_{1,2;{\uparrow},{\downarrow}}} = {\omega}\exp(i\pi\sqrt{\frac{E-E_0}{E_k}}),
\end{equation}
where $\omega$ is in energy unit and represents the strength of electrode-InSeI coupling, $E_k={\hbar}^2{\pi}^2/2ma^2$ is the characteristic kinetic energy of electrons in electrode with potential energy shift $E_0$. Typically we select $\omega = -0.8$ eV, $E_k=1.5$ eV and $E_0=-1.0$ eV (reference to InSeI CBM). Use Eq.~\ref{eq:H_uu}-\ref{eq:H_du} we obtain the Hamiltonian $H_C$ of an InSeI sample with $N$ unit cells as

\begin{equation}
    H_C = 
    \begin{bmatrix}
        H_0 & T & 0 & 0 & 0 & ... \\
        T^{\dagger} & H_1 & T & 0 & 0 & ... \\
        0 & T^{\dagger} & H_2 & T & 0 & ... \\
         & &  ...
    \end{bmatrix}_{N{\times}N}
\end{equation}

\[ 
    H_n = 
    \begin{bmatrix}
        h_{n,{\uparrow}{\uparrow}} & h_{{\uparrow}{\downarrow}} \\
        h_{{\downarrow}{\uparrow}} & h_{n,{\downarrow}{\downarrow}}
    \end{bmatrix}
\]

\[
    h_{n,{\uparrow}{\uparrow}} = 
    \begin{bmatrix}
        U_{1}+n{\Delta}{\phi} & i{\lambda_0} & 0 \\
        {\dagger} & U_{1}+n{\Delta}{\phi} & 0 \\
        {\dagger} & {\dagger} & U_{2}+n{\Delta}{\phi}
    \end{bmatrix}
\]

\[
    h_{n,{\downarrow}{\downarrow}} = 
    \begin{bmatrix}
        U_{1}+n{\Delta}{\phi} & -i{\lambda_0} & 0 \\
        {\dagger} & U_{1}+n{\Delta}{\phi} & 0 \\
        {\dagger} & {\dagger} & U_{2}+n{\Delta}{\phi}
    \end{bmatrix}
\]

\[ 
    h_{{\uparrow}{\downarrow}}=
    \begin{bmatrix}
        0 & 0 & {\lambda_R} \\
        0 & 0 & i{\lambda_R} \\
        -{\lambda_R} & -i{\lambda_R} & 0
    \end{bmatrix}
\]

\[ 
    T = 
    \begin{bmatrix}
        t_{{\uparrow}{\uparrow}} & 0 \\
        0 & t_{{\downarrow}{\downarrow}}
    \end{bmatrix}
\]

\[
    t_{{\uparrow}{\uparrow}} = 
    t_{{\downarrow}{\downarrow}} = 
    \begin{bmatrix}
        0 & it_1 & 0 \\
        -it_1 & 0 & 0 \\
        0 & 0 & t_2
    \end{bmatrix},
\]
where ${\Delta}\phi = {\Delta}V/N$ is the electrochemical potential bias per unit cell. 

\section{Classical interpretation to spin-selective transport in 1D helical crystal}\label{sec:classical details}

In the helical structure, any drifting motion of electron along $z$-direction will be accompanied by a rotational 
motion around the $z$-axis. In the rest reference system of the electron, this rotational motion causes the crystal electrostatic field to be time-dependent. We regard the crystal electrostatic potential energy $U$ as if it were due to an effective point charge in the distance $r$, such that $U=eV=eK/r=e|E|r$, where  $|E|=K/r^2$ is the effective electric field intensity and $V=K/r$ is the effective electric potential. According to the Maxwell equation, the time-dependent $E$ will induce a magnetic field $H$ as 
\begin{equation}
    {\nabla}{\times}\pmb{H}={\epsilon}{\partial}_t\pmb{E}
\end{equation}
or in the integral form 
\begin{equation}
    {\int}_{\Omega}{\nabla}{\times}\pmb{H}{\cdot}d\pmb{S}={\int}_{{\partial}{\Omega}}\pmb{H}{\cdot}d\pmb{l}={\int}_{\Omega}({\mu}{\epsilon}){\partial}_t\pmb{E}{\cdot}d\pmb{S}.
\end{equation}
Consider $\Omega$ as a rectangular region with length $dz$ in the $z$-direction and width $r$ in radius direction(${\perp}z$-direction), we obtain
\begin{equation}\label{eq:H_z}
     H_zdz=(\pm){\epsilon}|E|{\omega_e}rdz, 
\end{equation} 
where $\pm$ denotes the electron moving direction ${\pm}z$. Here, we assume that the rotating radius $r$ and the rotation angular frequency $\omega_e$ are constant (i.e. the rotation is circular), considering that each In-Se-I tetrahedra has the same structure. With the helix spatial period $a_0$ and electron velocity $v_z$, we have $\omega_e=2{\pi}/T=2{\pi}/(a_0/v_z)$. Considering $v_z$ is caused by external electric field $E_z$, we obtain $v_z=ME_z$, where $M$ is the mobility. Plugging these relations into equ-\ref{eq:H_z}, we have the magnetic induction intensity $B_z={H_z/{\mu}}$ as 
\begin{equation}
    |B_z| = \frac{\epsilon_r}{ec^2}{\omega_e}U = (\frac{1}{ec^2})(\frac{{\epsilon_r}UM}{a_0})E_z, 
\end{equation}
which will cause Zeeman energy splitting near CBM between spin-up and spin-down states as
\begin{equation}
    {\Delta}E_B = 2{\mu_B}B_z=(\frac{{\hbar}}{m_ec^2})\frac{{\epsilon_r}MU}{a_0}E_z, 
\end{equation}
where $\mu_B=e\hbar/2m_e$ is Bohr magneton. Consequently, the spin-up/down carrier density will be different, leading to a spin-polarized current 
\begin{equation}
    \begin{aligned}
        P &{\equiv} \frac{I_{\uparrow}-I_{\downarrow}}{I_{\uparrow}+I_{\downarrow}} = \frac{n_{\uparrow}-n_{\downarrow}}{n_{\uparrow}+n_{\downarrow}} \\
        & = \frac{1-\exp(-\frac{{\Delta}E_B}{kT})}{1+\exp(-\frac{{\Delta}E_B}{kT})} = \tanh(\frac{{\Delta}E_B}{2kT}) \\
        & = \tanh(E_z/E_{z0}),
    \end{aligned}
\end{equation}
via characteristic electric field intensity $E_{z0}$ as
\begin{equation}
    E_{z0} = (\frac{kT}{U})(\frac{m^*c^2}{\hbar})\frac{a_0}{{\epsilon_r}M}.
\end{equation} 
As an estimation, with room temperature $kT=0.026$ eV, total electrostatic potential $U \ {\sim} \ 15$ eV, $m_ec^2/\hbar=1.2{\times}10^{20}$ 1/s, $M=5.42{\times}10^{-3}{\rm m^2/(V{\cdot}s)}$, $a_0=30.0$\AA, ${\epsilon_r}{\sim}5$, the characteristic $E_{z0}{\sim}10^{10} \ {\rm V/m}$, which is around the order of magnitude of electric field intensity inside atom. 
Therefore, commonly a linear approximation is safe enough to be adopted as 
\[ P \ {\sim} \ E_z/E_{z0}. \]
It can be concluded that low temperature and strong external electric field are necessary conditions for observing a significant spin polarization. Any improvement to electron carrier mobility is also beneficial.

The hSOC can also be understood and estimated in a semi-classical way. Consider a quasi-electron with effective mass $m_e^*$ passing through the ions arranged along a helical chain with an axis towards $z$-direction, screw pitch $z_0$, and radius $r_0$. As the feature of helix, the electron should simultaneously have a linear motion with velocity $v_z= {\hbar}k/m_e^*$ and a rotational motion with angular velocity $\omega$. Once the electron covers a distance $z_0$ in $z$-direction, it should go through 2${\pi}$ in the circular direction, which means that they share the same period $T = z_0/v_z = 2{\pi}/\omega$ so that $\omega = 2\pi{\hbar}k/m_e^*z_0$. Recall that the classical OAM is $m_e^*\omega{r_0^2}$, we can calculate the OAM quantum number as
\begin{equation}
    \hbar{m'} = m_e^*{\omega}r_0^2 = \frac{2{\pi}{\hbar}r_0^2}{z_0}k,
\end{equation}
where $m' = 2k{\pi}r_0^2/z_0$ is proportional to $k$. In the rest reference framework of the electron,  the ions rotate relative to the electron, which induces a magnetic field with the magnetic induction strength along the $z$-direction as $B_z$.  The strength of the electrostatic interaction is denoted by the effective electric field intensity $E_0$ along the radius direction. We obtain $B_z \ {\sim} \ E_0\omega/2c^2$. Therefore, the spin magnetic moment $\mu_s = e{\hbar}s_z/2m_e^*$, with spin quantum number $s_z{\in}[-1/2, 1/2]$, is coupled with linear momentum $\hbar{k}$ (as long as the OAM $m'$) with the strength $\lambda_R$ as
\begin{equation}\label{eq:lambda_R}
    {\lambda_R}=\mu_sB_z=\frac{E_0{\hbar^2}{\pi}e}{2m_e^{*2}c^2z_0}ks_z = \frac{{\beta}}{z_0}ks_z.
\end{equation}


\bibliography{references}

\end{document}